\documentclass[10pt,aps,prx,twocolumn,superscriptaddress,longbibliography,reprint]{revtex4-2}

\usepackage[utf8]{inputenc}
\usepackage[T1]{fontenc}
\usepackage[english]{babel}
\usepackage{lmodern}

\usepackage{amsmath,amssymb,mathtools,bbm}
\usepackage{MnSymbol} 

\mathtoolsset{
  above-shortintertext-sep=-.5\belowdisplayshortskip, 
  below-shortintertext-sep=-.5\abovedisplayskip, 
  above-intertext-sep=-.25\belowdisplayskip, 
  below-intertext-sep=-.25\abovedisplayskip, 
}

\usepackage[export]{adjustbox} 
\usepackage[shortlabels,inline]{enumitem}
\usepackage{booktabs,makecell}

\usepackage{silence}
\usepackage[colorlinks=true,allcolors=blue]{hyperref}

\newcommand{\dd}[0]{\mathrm d}
\newcommand{\dD}[0]{\mathrm D}
\newcommand{\ee}[0]{\mathrm e}
\newcommand{\ii}[0]{\mathrm i}

\DeclareMathOperator{\tr}{tr}
\DeclareMathOperator{\Res}{Res}
\DeclarePairedDelimiter{\abs}{\lvert}{\rvert}
\DeclarePairedDelimiter{\norm}{\lVert}{\rVert}
\DeclarePairedDelimiter{\expval}{\langle}{\rangle}
\DeclarePairedDelimiter{\cumulant}{\llangle}{\rrangle}
\DeclarePairedDelimiterX{\comm}[2]{[}{]}{#1, #2}
\DeclarePairedDelimiterX{\acomm}[2]{\{}{\}}{#1, #2}

\DeclarePairedDelimiter{\ket}{\lvert}{\rangle}
\DeclarePairedDelimiterX{\braket}[2]{\langle}{\rangle}{#1\!\delimsize\mid\mathopen{}\!#2}
\DeclarePairedDelimiterX{\ketbra}[2]{\vert}{\vert}{#1\,\delimsize\rangle\!\delimsize\langle\,\mathopen{}#2}

\newcommand{\op}[1]{\mathsf #1}

\newcommand{\up}[0]{\mathrm{u}}
\newcommand{\dwn}[0]{\mathrm{d}}

\begin{document}

\title{Full counting statistics and first-passage times in quantum Markovian processes:\texorpdfstring{\\}{} Ensemble relations, metastability, and fluctuation theorems}

\author{Paul Menczel}
\email{paul@menczel.net}
\affiliation{RIKEN Center for Quantum Computing, RIKEN, Wakoshi, Saitama 351-0198, Japan}

\author{Christian Flindt}
\affiliation{Department of Applied Physics, Aalto University, 00076 Aalto, Finland}
\affiliation{RIKEN Center for Quantum Computing, RIKEN, Wakoshi, Saitama 351-0198, Japan}

\author{Fredrik Brange}
\affiliation{Department of Applied Physics, Aalto University, 00076 Aalto, Finland}

\author{Franco Nori}
\affiliation{RIKEN Center for Quantum Computing, RIKEN, Wakoshi, Saitama 351-0198, Japan}
\affiliation{Physics Department, University of Michigan, Ann Arbor, MI 48109-1040, USA}

\author{Clemens Gneiting}
\affiliation{RIKEN Center for Quantum Computing, RIKEN, Wakoshi, Saitama 351-0198, Japan}

\date{\today}

\begin{abstract}
We develop a comprehensive framework for characterizing fluctuations in quantum transport and nonequilibrium thermodynamics using two complementary approaches: full counting statistics and first-passage times.
Focusing on open quantum systems governed by Markovian Lindblad dynamics, we derive general ensemble relations that connect the two approaches at all times, and we clarify how the steady states reached at long times relate to those reached at large jump counts.
In regimes of metastability, long-lived intermediate states cause violations of experimentally testable cumulant relations, as we discuss.
We also formulate a fluctuation theorem governing the probability of rare fluctuations in the first-passage time distributions based on results from full counting statistics.
Our results apply to general integer-valued trajectory observables that do not necessarily increase monotonically in time.
Three illustrative applications, a two-state emitter, a driven qubit, and a variant of the Su-Schrieffer-Heeger model, highlight the physical implications of our results and provide guidelines for practical calculations.
Our framework provides a complete picture of first-passage time statistics in Markovian quantum systems, encompassing multiple earlier results, and it has direct implications for current experiments in quantum optics, superconducting circuits, and nanoscale heat engines.
\end{abstract}

\maketitle

\section{Introduction}

\begin{figure}[b!]%
	\centering%
	\vspace{-1.2em}%
	\includegraphics[scale=1]{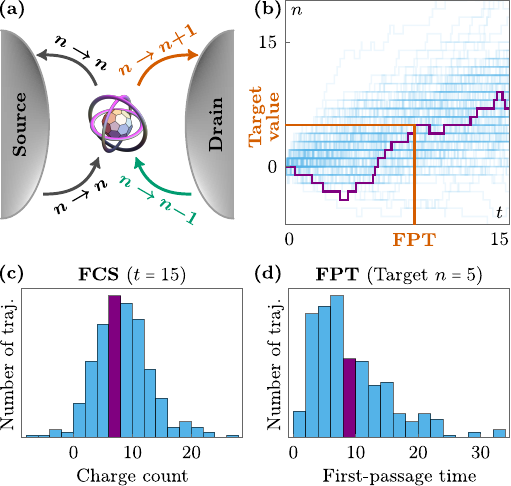}%
	\caption[]{
        Full counting statistics and first-passage times.
        \begin{enumerate*}[(a)]
		\item As an example, we consider particles that tunnel between an open quantum system and  two external leads.
            The tunneling events to and from the drain are detected and change the number of transferred particles~$n$ by plus or minus one.
		\item The resulting stochastic trajectories show the number of transferred particles as a function of time.
            The first-passage time (FPT) is the first time that a given target value of $n$ is reached.
		\item The full counting statistics~(FCS), by contrast, is the distribution of the particle number $n$ at a fixed time $t$, here $t=15$ (a.u.).
		\item The first-passage time distribution for the target value $n=5$.
		\end{enumerate*}
		The highlighted example trajectory in panel~(b) falls into the highlighted bins in the two histograms.
	}%
	\label{fig:1}%
\end{figure}

\begin{figure*}[t]
	\centering
    \includegraphics[scale=1]{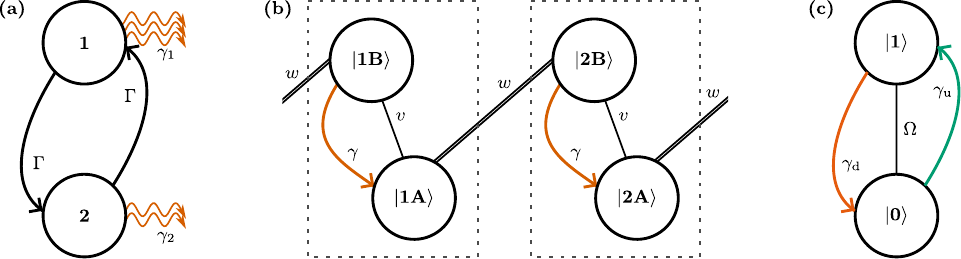}
	\caption[]{
		Applications of full counting statistics and first-passage times.
		\begin{enumerate*}[(a)]
		\item Photon emitter, which switches stochastically between two states at the rate $\Gamma$.
            Photons are emitted at the rates $\gamma_1$ or $\gamma_2$, depending on the state of the system.
            This setup is discussed in Sec.~\ref{subsec:exampleTWE}.
		\item Su-Schrieffer-Heeger (SSH) model, where each unit cell $i$ contains two sites, $\ket{iA}$ and $\ket{iB}$.
            The tunneling amplitude between sites within a cell is denoted by~$v$, while $w$ connects neighboring cells.
            We consider two unit cells and impose periodic boundary conditions.
            The orange arrows represent a collective dissipation channel, which induces quantum jumps from $\ket{iB}$ to $\ket{iA}$.
            We discuss the SSH model in Sec.~\ref{sec:exampleSSH}.
		\item Driven qubit, where transitions between two states are driven coherently with the Rabi frequency $\Omega$ and incoherently at the rates $\gamma_\dwn$ and~$\gamma_\up$.
            We consider this setup in Sec.~\ref{subsec:example3}.
		\end{enumerate*}
	}
	\label{fig:examples}
\end{figure*}

Fluctuations are a defining feature of quantum systems operating far from equilibrium~\cite{HanggiPhysRep1982, Gardiner2004, vanKampen2011}.
Systems with applications in quantum technology include superconducting circuits~\cite{DevoretScience2013, PekolaNatPhys2015, GuPhysRep2017, KockumFundamentalsandFrontiersoftheJosephsonEffect2019}, quantum dots~\cite{KastnerRevModPhys1992, KoskiNatPhys2013, GustavssonPhysRevLett2006}, and nanoscale heat engines~\cite{RossnagelScience2016, JosefssonNatNanotechnol2018, KlatzowPhysRevLett2019, PetersonPhysRevLett2019, GuthriePhysRevAppl2022}, where fluctuations are a fundamental part of their working principle.
Characterizing these fluctuations is essential for understanding a wide range of physical processes in fields such as quantum transport~\cite{Dittrich1998, Nazarov2009}, quantum control~\cite{DAlessandro2007, ZhangPhysRep2017}, and quantum thermodynamics~\cite{Binder2018, Kurizki2022}.
A particular type of fluctuations, which is of interest here, relates to the random quantum jumps that occur in open quantum systems as they exchange heat or particles with an environment~\cite{PlenioRevModPhys1998, GneitingPhysRevA2021, LandiPRXQuantum2024}.

As illustrated in Fig.~\ref{fig:1}(a), the quantum jumps may correspond to the tunneling of single electrons between a central quantum system and one or more external electrodes~\cite{SinghPhysRevLett2019}.
One may then monitor the number of electrons~$n$ that have been collected in one of them and map out trajectories of the electron number as a function of time as shown in Fig.~\ref{fig:1}(b), where each trajectory corresponds to a single experimental realization.
There are now several ways to characterize such an ensemble of trajectories.
On the one hand, one may consider the full counting statistics, where one builds a histogram of the electron number at a fixed time $t$, as shown in Fig.~\ref{fig:1}(c), to obtain the distribution $P_t(n)$ for this random variable~\cite{BlanterPhysRep2000, Nazarov2003, LandiPRXQuantum2024}.
In general, full counting statistics describes the probability distribution of time-integrated observables, such as the total transferred charge or accumulated energy within a given time interval.
Full counting statistics has been used to characterize quantum transport, and it is closely related to large-deviation theory~\cite{TouchettePhysRep2009} and fluctuation theorems~\cite{JarzynskiPhysRevLett1997, CrooksPhysRevE1999, SeifertPhysRevLett2005, EspositoRevModPhys2009, CampisiRevModPhys2011}. 
On the other hand, one can fix a target value for the electron number~$n$ and ask how long it takes before this value is reached for the first time.
One can then build a histogram of the first-passage time, as illustrated in Fig.~\ref{fig:1}(d), to obtain the first-passage time distribution~$P_n(t)$.
In this way, first-passage times offer a time-resolved view of dynamical processes.
For an overview of first-passage time distributions, we refer to the books in Refs.~\cite{Redner2001, Metzler2014}, and in the context of open quantum systems, they have recently been investigated in Refs.~\cite{ManzanoPhysRevLett2019, RudgePhysRevB2019, RudgePhysRevB2019a, RudgeJChemPhys2019, VanVuPhysRevLett2022, HasegawaPhysRevE2022, HePhysRevAppl2023, KewmingPhysRevA2024, LiuPhysRevE2024, Bakewell-SmithJStatPhys2025}.

While full counting statistics captures the accumulated behavior, first-passage time distributions describe the temporal structure of rare events.
Although the two approaches appear to be conceptually different, they are also related in several ways.
In this work, we rigorously establish general ensemble relations between the two descriptions, building on earlier work that identified a connection between them in the limit of long observation times and large jump counts for one-way currents, where the counted quantity can only increase~\cite{BudiniJStatMech2014}.
Specifically, it was found that the probability of measuring the average output current $j$ over a long observation time relates to the probability for the average first-passage time to have the value $\tau = 1 / j$ for large target counts.
These results have then been generalized to two-way currents~\cite{GingrichPhysRevLett2017}, where the accumulation of particles is non-monotonic and the target value can be reached several times as seen in Fig.~\ref{fig:1}(b).
With that extension, it is possible to reformulate thermodynamic uncertainty relations~\cite{BaratoPhysRevLett2015, GingrichPhysRevLett2016, SeifertPhysicaA2018, HorowitzNatPhys2020} for the full counting statistics as uncertainty relations for the first-passage time distributions, which have implications for the performance of quantum clocks~\cite{MeierNatPhys2025, MacieszczakArXiv240709839Cond-Matstat-Mech2024}.

\subsection*{Summary of This Work}

In this paper, we employ a rigorous theoretical framework for open quantum systems described by Lindblad master equations~\cite{GoriniJMathPhys1976, LindbladCommunMathPhys1976, Breuer2002, LidarArXiv190200967Quant-Ph2020} to address three main objectives.
First, we generalize the asymptotic correspondences from earlier works to arbitrary times by deriving general relations between the dynamical generators of the full counting statistics and the first-passage time distributions.
The generator of the full counting statistics describes the evolution of the wall-time ensemble, which is the ensemble-averaged state at a fixed time as shown by a clock on the wall.
The generator of the first-passage time distribution governs the evolution of the jump-time ensemble, which is the ensemble average of the system state at the first-passage time~\cite{GneitingPhysRevA2021}.
Our relation between the generators provides a practical computational tool, and it sheds light on the connections between the steady states and dynamical observables in the two ensembles.

Our second objective is to identify the conditions under which the asymptotic correspondences and the ensemble relations hold and when they may break down.
To this end, our theoretical framework allows us to clarify the situations where the correspondence between the generators or the asymptotic correspondences may not hold.
In particular, we show that a violation of certain relations between the cumulants of the two distributions indicates the presence of metastable states~\cite{MacieszczakPhysRevLett2016, MacieszczakPhysRevRes2021}, and it can therefore be used to detect regimes of metastability.

Our third objective is to investigate the relationship between the first-passage time distributions for positive and negative target values.
Specifically, we derive a fluctuation theorem that quantifies how rare it is for trajectories to reach large negative target values.
We also show how the exponent of this fluctuation theorem relates to spectral properties of the generator of the full counting statistics, which again connects the two ensembles. 

By developing a unified approach to time-integrated and time-resolved fluctuations, our work provides a theoretical foundation for investigating nonequilibrium dynamics of open quantum systems.
To illustrate our findings, we consider the three systems in Fig.~\ref{fig:examples}, showing a two-state emitter, the Su-Schrieffer-Heeger (SSH) model~\cite{SuPhysRevLett1979, GneitingPhysRevRes2022}, and a driven quantum two-level system.
We focus here on Markovian open quantum systems with discrete quantum jumps.
However, our methodology may also find use in other fields, for example in classical stochastic dynamics~\cite{Gardiner2010, GoldhirschPhysRevA1986, GoldhirschPhysRevA1987, RoldanPhysRevLett2015, SaitoEPL2016, NeriPhysRevX2017, ArtimePhysRevE2018, NeriPhysRevLett2020, ManzanoPhysRevLett2021}, astrophysics~\cite{ChandrasekharAstrophysJ1943}, biochemistry~\cite{BergBiophysJ1977, SzaboJChemPhys1980, ZwanzigProcNatlAcadSci1992, BenichouNatChem2010}, computer science~\cite{MajumdarTheLegacyofAlbertEinstein2006}, or finance~\cite{PerelloPhysRevE2011}.

Our article is divided into two parts.
In Sec.~\ref{sec:counting}, we consider one-way currents, where the number of counted particles can only increase.
This assumption simplifies the theory of first-passage times, since a given target value can only be reached once.
In Sec.~\ref{sec:bidi}, we then consider two-way currents, where the number of transferred particles can both increase and decrease, and we extend our framework for one-way currents to two-way currents.

The first part of our article is organized as follows.
In Sec.~\ref{subsec:uni:setup}, we describe the  physical setups that we consider together with our theoretical framework based on the Lindblad master equation.
In Secs.~\ref{subsec:uni:fcs} and \ref{subsec:uni:fpt}, we introduce the full counting statistics and the first-passage time distribution for one-way currents and relate it to the Lindblad master equation.
In Sec.~\ref{subsec:main_thm}, we show how the full counting statistics and the first-passage time distribution are related, extending the earlier results from Refs.~\cite{BudiniJStatMech2014,GingrichPhysRevLett2017}.
In Sec.~\ref{subsec:steady_states}, we discuss steady states in the thermodynamic limit of long times and many quantum jumps as well as the relation between observables measured in these steady states.
In Sec.~\ref{subsec:cumulant_relations}, we recover a correspondence between the generating functions of the full counting statistics and of the first-passage time distribution, and we identify the conditions under which it can be violated.
In Secs.~\ref{subsec:exampleTWE} and~\ref{sec:exampleSSH}, we illustrate many of these concepts with a model of a two-state photon emitter and with the Su-Schrieffer-Heeger (SSH) model.

The second part of our article is organized as follows.
In Sec.~\ref{subsec:bidi:setup}, we extend our theoretical framework to include two-way currents.
In Sec.~\ref{subsec:bidi:fcs}, we show that this extension is relatively straightforward for the full counting statistics. In Sec.~\ref{subsec:bidi:fpt}, by contrast, we find that the extension is much more involved for the first-passage time distribution.
Still, in Sec.~\ref{subsec:bidi:correspondence}, we demonstrate that the correspondence between the full counting statistics and the first-passage time distribution carries over to this more general setting.
Moreover, for two-way currents, we show in Sec.~\ref{subsec:fluctuation_theorem} how fluctuation theorems for the full counting statistics can be reformulated as fluctuation theorems for the first-passage time distribution.
In Sec.~\ref{subsec:example3}, we illustrate several of these findings with a model of a coherently driven and dissipative qubit.

Finally, we conclude in Sec.~\ref{sec:perspectives} and provide perspectives for future research.
Several technical details of our work are presented in the appendices. 
For convenience, we provide an alphabetical list of commonly used symbols and notations in Table~\ref{tab:notation} in the Appendix.

\section{One-way currents} \label{sec:counting}

\subsection{Setup} \label{subsec:uni:setup}

For our general theory, we consider a quantum system as depicted in Fig.~\ref{fig:1}(a), where a quantum dot as an example makes transitions between different charge states as electrons tunnel to and from two external electrodes.
However, in the following, it should be clear that our theory applies to any open quantum system that makes discrete transitions between different quantum states.

The density matrix of the quantum system, $\rho_t$, evolves according to the Markovian master equation
\begin{equation} \label{eq:master_equation}
	\partial_t \rho_t = \op L \rho_t,
\end{equation}
where the superoperator $\op L$ is the generator of the time evolution.
It has the Lindblad form~\cite{GoriniJMathPhys1976, LindbladCommunMathPhys1976, Breuer2002, LidarArXiv190200967Quant-Ph2020}
\begin{equation} \label{eq:lindblad}
	\op L \rho = -\ii \comm{H}{\rho} + \sum_\mu \gamma_\mu\, \Bigl[ L_\mu \rho L_\mu^\dagger - \frac 1 2 \{L_\mu^\dagger L_\mu, \rho\} \Bigr],
\end{equation}
where $H$ is the Hamiltonian of the system and $\gamma_\mu > 0$ are the relaxation rates for the dissipation channels described by the Lindblad operators $L_\mu$.
The curly brackets denote the anti-commutator, and we take $\hbar=1$ for the reduced Planck's constant.
For the sake of clarity, we focus on finite-dimensional systems, but we expect that our findings also  hold for physical systems of infinite dimensions.

Our general formulation encompasses classical systems, where the master equation \eqref{eq:master_equation} is a rate equation for stochastic transitions between a discrete set of states~\cite{SchnakenbergRevModPhys1976, SeifertRepProgPhys2012}. The density matrix is then diagonal, and we can write it as a probability vector whose $i$-th entry $(\rho_t)_i$ is the probability of finding the system in the state $i$ at the given time.
The generator $\op L$ is then a rate matrix
\begin{equation} \label{eq:classical_ME}
	\op L_{ij} = \gamma_{ij} - \delta_{ij} \sum_k \gamma_{kj},
\end{equation}
where $\gamma_{ij}$ is the transition rate from state $j$ to state $i$.

In both the classical and the quantum case, the master equation describes the evolution of an ensemble of realizations of an experiment.
Although the ensemble average $\rho_t$ evolves smoothly, the individual realizations are characterized by sudden transitions between different system states.
For quantum systems, the transitions are induced by the dissipation channels and are referred to as quantum jumps~\cite{DalibardPhysRevLett1992, MolmerJOptSocAmB1993, PlenioRevModPhys1998}.
The quantum jumps may correspond to the exchange of particles, energy or charge between the quantum system and its environment~\cite{EspositoRevModPhys2009, LandiPRXQuantum2024}, and our aim is to describe these exchange processes statistically.

In the first part of this paper, we focus on one-way currents.
For our example in Fig.~\ref{fig:1}(a), that means that particles can only be emitted into the drain electrode but not tunnel back from the drain.
Therefore, jump events either increase the number of counted particles [orange arrow in Fig.~\ref{fig:1}(a)] or are ignored (gray arrows), but for now we exclude the possibility of jump events that would decrease the particle count (green arrow).
In the second part of the paper, we consider the more general case where the counter can both increase and decrease.

We specify the types of events that are counted by splitting the generator into two contributions~\cite{FlindtEPL2004}
\begin{equation} \label{eq:jump_op}
	\op L = \op L_0 + \op J,
\end{equation}
where $\op J$ is the jump operator for the transitions that we count. The operator $\op L_0 = \op L - \op J$ is the generator of the time evolution without these transitions, but it may include other transitions that we ignore.

For classical systems, the jump operator is a matrix whose entry in the $i$-th row and $j$-th column is $\gamma_{ij}$ if the transition from state~$j$ to state $i$ is among the ones that we count, and zero otherwise.
For quantum systems, it is defined as $\op J \rho = \sum_\mu^\prime \gamma_\mu L_\mu \rho L_\mu^\dagger$, where the prime indicates that the sum runs over only those dissipation channels that are counted.
For imperfect detection, one may multiply each term in $\op J$ by the corresponding detection probability.

Each realization of the experiment can now be characterized by the jump count $n$ and the times $0 \leq t_1 \leq t_2 \leq \cdots$ at which the corresponding transitions occurred.
This jump count does not include jump events that are ignored.
Given an initial state $\rho_0$ and a time span $[0,t]$, the probability density for obtaining the jump record $(t_1\,\cdots\,t_n)$ reads~\cite{Carmichael1993, MenczelPhysRevRes2020}
\begin{equation} \label{eq:trajectory_weight}
	P_t(t_1 \,\cdots\, t_n) = \tr\Bigl[ \ee^{\op L_0 (t - t_n)}\, \op J\, \ee^{\op L_0 (t_n - t_{n-1})}\, \cdots\, \op J\, \ee^{\op L_0 t_1} \rho_0 \Bigr].
\end{equation}
To stress that this probability density does not allow for additional jumps to be counted outside of the given times $t_k$, it is also called the \emph{exclusive} probability density \cite{Carmichael1993}.
Moreover, the state at the final time, conditioned on the jump record, is
\begin{equation} \label{eq:trajectory}
	\rho_{t\, \mid\,  t_1 \,\cdots\, t_n} = \frac{1}{P_t(t_1 \,\cdots\, t_n)}\, \ee^{\op L_0 (t - t_n)}\, \op J\, \ee^{\op L_0 (t_n - t_{n-1})}\, \cdots\, \mathsf J\, \ee^{\op L_0 t_1} \rho_0.
\end{equation}
These conditioned states, viewed as functions of the final time, are called trajectories.
In our formalism, they are not necessarily pure states.
By averaging over all trajectories, we recover the full ensemble as
\begin{equation} \label{eq:all_trajectories}
	\rho_t = \sum_{n=0}^\infty \int_0^t \dd t_1\, \cdots\, \int_{t_{n-1}}^t \dd t_n\, P_t(t_1 \,\cdots\, t_n)\, \rho_{t\, \mid\, t_1 \,\cdots\, t_n},
\end{equation}
which solves the master equation \eqref{eq:master_equation}.

\subsection{Full Counting Statistics} \label{subsec:uni:fcs}

The full counting statistics is the statistical distribution of the jump count at a fixed time $t$~\cite{LevitovJMathPhys1996, BagretsPhysRevB2003, FlindtEPL2004, LandiPRXQuantum2024}. An example is shown in Fig.~\ref{fig:1}(c). To find the full counting statistics, we divide the ensemble \eqref{eq:all_trajectories} into partitions according to the number of jumps along the trajectories,
\begin{equation} \label{eq:Tn_trajectories}
	\rho_t(n) = \int_0^t \dd t_1\, \cdots\, \int_{t_{n-1}}^t \dd t_n\, P_t(t_1 \,\cdots\, t_n)\, \rho_{t\, \mid\, t_1 \,\cdots\, t_n},
\end{equation}
such that $\rho_t = \sum_{n=0}^\infty \rho_t(n)$.
The operators $\op T_t(n)$ that map the initial state $\rho_0$ to the ensemble partitions at time $t$, $\rho_t(n) = \op T_t(n) \rho_0$, are called the $n$-resolved propagators.
They evolve according to the differential equation
\begin{equation} \label{eq:n-resolved-prop}
	\partial_t \op T_t(n) = \op L_0 \op T_t(n) + \op J \op T_t(n-1)
\end{equation}
with the initial condition $\op T_0(n) = \delta_{n,0}$ and $\op T_t(-1) = 0$.

By definition, the quantity
\begin{equation} \label{eq:fcs}
	P_t(n) = \tr \rho_t(n)
\end{equation}
is the probability that the jump count has the value $n$ at the time $t$.
Under the condition that the jump count is $n$, the (unnormalized) system state at that time is $\rho_t(n)$.

The probabilities $P_t(n)$ form the full counting statistics. To define the cumulants of this probability distribution, it is useful to introduce the generating function
\begin{equation} \label{eq:fcs-gf}
	c_t(z) = \ln \sum_n z^{-n} P_t(n).
\end{equation}
For the sake of notational convenience later on, our definition differs from the conventional cumulant-generating function, which would be obtained by replacing $z$ by $e^{-x}$ in the expression above.
Note also that $c_t(1) = 0$, since the probabilities sum to one.
With our definition, the cumulants are extracted from the generating function as
\begin{equation} \label{eq:fcs-cumulants}
	\cumulant{n^k}_t = \partial_x^k c_t(\ee^{-x}) \bigr|_{x=0}.
\end{equation}
The first cumulant is the average jump count at time $t$, $\cumulant{n}_t = \expval{n}_t$, while the second cumulant is the variance of the jump count, $\cumulant{n^2}_t = \expval{n^2}_t - \expval{n}_t^2$, which is given by the first and second moments~\cite{vanKampen2011}.
As the third cumulant and all higher cumulants vanish for Gaussian distributions, these cumulants quantify deviations from Gaussianity.

Using the definition \eqref{eq:fcs}, the generating function can be rewritten as 
\begin{equation}
	c_t(z) = \ln \tr \rho_t(z),
\end{equation}
where $\rho_t(z) = \sum_n z^{-n} \rho_t(n)$ is the state including the counting field $z$.
The equation of motion for the state with the counting field has the solution 
\begin{equation} \label{eq:rho_with_cf}
	\rho_t(z) = \exp\bigl[ \op L(z) t \bigr]\, \rho_0 ,
\end{equation}
where
\begin{equation} \label{eq:tilted_gen}
	\op L(z) = \op L_0 + z^{-1} \op J
\end{equation}
is referred to as the tilted generator~\cite{BagretsPhysRevB2003, FlindtEPL2004}.
For $z > 1$, it describes the time evolution of a system that is subject to post-selection~\cite{MingantiPhysRevA2020, GuArXiv250306946Quant-Ph2025}.
Inserting Eq.~\eqref{eq:rho_with_cf} into Eq.~\eqref{eq:fcs-gf}, we arrive at the expression
\begin{equation} \label{eq:fcs-gf2}
	c_t(z) = \ln \tr\bigl( \exp\bigl[\op L(z) t \bigr]\, \rho_0 \bigr)
\end{equation}
for the generating function.
If $\op L(z)$ has a unique eigenvalue with maximum real part (ignoring multiplicities), the generating function grows linearly with $t$ at long times, and the scaled generating function is then given by this particular eigenvalue as~\cite{BagretsPhysRevB2003}
\begin{equation} \label{eq:fcs-sgf}
	c(z) = \lim_{t \to \infty} \frac 1 t\, c_t(z) = \lambda_{\mathrm{max}}^{\mathrm{re}}[ \op L(z) ].
\end{equation}

In the following, we assume that the system cannot sustain coherent oscillations at long times, i.e., that the dynamical generator $\op L$ does not have any purely imaginary eigenvalues.
The scaled generating function is then well-defined in a neighborhood of $z = 1$.
If, in addition, it is a smooth function in that neighborhood, then the cumulants also grow linearly at long times, and we can define
the scaled cumulants as
\begin{equation} \label{eq:fcs_cumulants}
	\cumulant{j^k}=\lim_{t \to \infty} \frac 1 t\, \cumulant{n^k}_t = \partial_x^k c(\ee^{-x}) \bigr|_{x=0} .
\end{equation}
If there is an eigenvalue crossing at $z=1$, the scaled generating function is not smooth, and the cumulants may grow faster than linearly at long times.
We will return to this aspect in Sec.~\ref{subsec:cumulant_relations}.

\subsection{First-Passage Times} \label{subsec:uni:fpt}

We now go on to discuss the first-passage time statistics, which is the statistical distribution of the first time that the jump count reaches a given target value~$n$~\cite{Redner2001}.
An example of a first-passage time distribution is shown in Fig.~\ref{fig:1}(d). 
The ensemble of trajectories that reaches the target value at the time $t$ reads
\begin{equation}
\label{eq:barrho_n}
	\bar\rho_n(t) = \int_0^t \dd t_1\, \cdots\, \int_{t_{n-1}}^t \dd t_n\, \delta(t - t_n)\, P_t(t_1\, \cdots\, t_n)\, \rho_{t\, \mid\, t_1\,\cdots\, t_n}.
\end{equation}
We then define the first-passage propagators $\op F_n(t)$ as the operators that map the initial state to the ensembles $\bar\rho_n(t) = \op F_n(t) \rho_0$.
A comparison with Eq.~\eqref{eq:Tn_trajectories} shows that they are given by
\begin{equation} \label{eq:fnt}
	\op F_n(t) = \op J \op T_t(n-1).
\end{equation}

The first-passage time distribution reads
\begin{equation} \label{eq:fptd}
	\bar P_n(t) = \tr \bar\rho_n(t) .
\end{equation}
The (unnormalized) system state right after the $n$-th jump, conditioned on the first-passage time being $t$, is given by $\bar\rho_n(t)$. To confirm that this definition of the first-passage time distribution agrees with our basic intuition, we may additionally consider the relation
\begin{equation} \label{eq:fptd_relation}
	\bar P_n(t) = \partial_t \sum_{k \geq n} P_t(k),
\end{equation}
which holds only in the case of one-way currents, and can be derived from the definitions by using that $\op L$ is trace-preserving.
This relation shows that, as one might expect, the probability of having a count of $n$ or larger at time $t$ is the cumulative distribution function for the first-passage time distribution.

Using this relation, the probability of never detecting any quantum jumps becomes
\begin{equation}
	\lim_{t\to\infty} P_t(0) = 1 - \lim_{t\to\infty} \sum_{k \geq 1} P_t(k) = 1 - \int_0^\infty \dd t\, \bar P_1(t),
\end{equation}
which is the complement of the integrated first-passage time distribution. The probability $P_t(0) = \tr[ \ee^{\op L_0 t} \rho_0 ]$ vanishes at long times, independently of the initial condition, if the operator $\op L_0$ has no zero eigenvalues.
In Appendix~\ref{app:dark_states}, we show that zero eigenvalues of $\op L_0$ correspond to dark states; therefore, the first-passage time distribution is normalized if and only if there are no dark states.
Dark states, or absorbing states, are states that trajectories are unable to leave.
Quantum phase transitions involving dark states were recently explored in Refs.~\cite{MarcuzziPhysRevLett2016, CarolloPhysRevLett2019a, CarolloPhysRevLett2022, PerfettoPhysRevLett2023}.

In analogy to our discussion of the full counting statistics, we now introduce the moment-generating function
\begin{equation} \label{eq:fps-gf}
	\bar c_n(s) = \int_0^\infty \dd t\, \ee^{-st}\, \bar P_n(t)
\end{equation}
for the first-passage time distribution \footnote{
Formally, the integral is only convergent for parameters $s$ with large enough real parts. However, the functions $\bar c_n(s)$ can be analytically continued to holomorphic functions on the entire complex plane, except for some poles and branch cuts. Here, we always implicitly consider the analytic continuations of conditionally convergent integrals such as this one.
}.
In this case, the definition of the moment-generating function agrees with the conventional one for continuous random variables, which is the Laplace transform of the distribution.
The cumulants of the first-passage time are then given as the derivatives
\begin{equation} \label{eq:fpt-cumulants}
	\cumulant{t^k}_n = (-\partial_s)^k \ln\bigl[ \bar c_n(s) \bigr] \bigr|_{s=0}.
\end{equation}

\begin{table*}[t]
	\centering
	\renewcommand{\arraystretch}{1.35}
	\setlength\tabcolsep{0pt}
	\begin{tabular*}{\linewidth}{@{\extracolsep{\fill}}llll@{}}
		\toprule
		\bf Quantity & \bf Full Counting Statistics & \bf First-Passage Times & \bf Correspondence \\
		\midrule
		\multicolumn{4}{c}{\textsc{Basic Definitions (Secs.~\ref{subsec:uni:fcs} and \ref{subsec:uni:fpt})}} \\
		\midrule
		Independent variable & Time $t$ & Jump count $n$ \\
		Dependent variable & Jump count $n$ & Time $t$ \\
		-- its conjugate & Counting field $z$ & Laplace parameter $s$ \\
		System state & Wall-time average $\rho_t$ & Jump-time average $\bar\rho_n$ \\ \addlinespace[9pt]
		\midrule
		\multicolumn{4}{c}{\textsc{Finite-Time Correspondence (Sec.~\ref{subsec:main_thm})}} \\
		\midrule
		Dynamical generator & Tilted generator $\op L(z)$ & First-passage prop.\ $\op F_1(s)$ & Expressed in terms of eigensystems: \\
		-- its eigenvalues (EV) & $s$ with $\det[\op L(z)-s]=0$ & $z$ with $\det[\op F_1(s)-z]=0$ & \makecell[tl]{$\op L(z)$ has EV $s$ $\Leftrightarrow$ $\op F_1(s)$ has EV $z$\\ (exceptions discussed in Sec.~\ref{subsec:main_thm})} \\
		-- its right eigenstates & $\rho$ with $\op L(z)\, \rho = s\rho$ & $\bar\rho$ with $\op F_1(s)\, \bar\rho = z \bar\rho$ & $\rho = \op T_s(0)\, \bar\rho$ \\
		-- its left eigenstates & $\varphi$ with $\varphi\, \op L(z) = s\varphi$ & $\bar\varphi$ with $\bar\varphi\, \op F_1(s) = z \bar\varphi$ & $\varphi = \bar\varphi$ \\
		\multicolumn{4}{l}{This correspondence can be used, for example, to compute $\op F_1(s)$ from eigensystem of $\op L(z)$, see Eq.~\eqref{eq:uni:F_explicit}.} \\ \addlinespace[9pt]
		\midrule
		\multicolumn{4}{c}{\textsc{Steady State Correspondence (Sec.~\ref{subsec:steady_states})}} \\
		\midrule
		Evolution equation & $\rho_t = \exp(\op L t)\, \rho_0$ with $\op L = \op L(1)$ & $\bar\rho_n = \op F_1^n\, \rho_0$ with $\op F_1 = \op F_1(0)$ \\
		Steady states & \makecell[tl]{Wall-time steady state:\\ $\rho$ with $\op L\, \rho = 0$\; (wtss)} & \makecell[tl]{Jump-time steady state:\\ $\bar\rho$ with $\op F_1\, \bar\rho = \bar\rho$\; (jtss)} & \makecell[tl]{$\rho$ is wtss $\Rightarrow$ $\op J\, \rho$ is jtss\\ $\bar\rho$ is jtss $\Rightarrow$ $\op L_0^{-1}\, \bar\rho$ is wtss} \\
		Dynamical steady states & $\rho_\infty = \lim_{t\to\infty} \rho_t$ & $\bar\rho_\infty = \lim_{n\to\infty} \bar\rho_n$ & \makecell[tl]{$\bar\rho_\infty \propto \op J\, \rho_\infty$ and $\rho_\infty \propto \op L_0^{-1}\, \bar\rho_\infty$\\ (if steady state is unique)} \\
		Observables & $\expval{A}_{\text{wt}} = \tr[A \rho_\infty]$ & $\expval{\bar A}_{\text{jt}} = \tr[\bar A \bar\rho_\infty]$ & \makecell[tl]{$\expval{A}_{\text{wt}} = \expval{\bar A}_{\text{jt}}$ for $\bar A = -j_{\text{wt}}\, (\op L_0^\dagger)^{-1} A$\\ (if steady state is unique)} \\ \addlinespace[9pt]
		\midrule
		\multicolumn{4}{c}{\textsc{Cumulant Correspondence (Sec.~\ref{subsec:cumulant_relations})}} \\
		\midrule
		Generating function & $c_t(z) = \tr\{ \exp[\op L(z) t]\, \rho_0 \}$ & $\bar c_n(s) = \tr[ \op F_1(s)^n\, \rho_0 ]$ \\
		Scaled generating fct. & $c(z) = \lim_{t\to\infty} c_t(z) / t$ & $\bar c(s) = \lim_{n\to\infty} \bar c_n(s)^{1/n}$ & \makecell[tl]{$c(\bar c(s)) = s$ and $\bar c(c(z)) = z$\\ (derived in Ref.~\cite{BudiniJStatMech2014})} \\
		Scaled cumulants & $\cumulant{j^k} = \partial_x^k c(\ee^{-x}) \bigr|_{x=0}$ & $\cumulant{\tau^k} = (-\partial_s)^k \ln\bigl[ \bar c(s) \bigr] \bigr|_{s=0}$ & \makecell[tl]{Follows from the row above, e.g.:\\$\cumulant{j} = 1/\cumulant{\tau}$ and $\cumulant{j^2} = \cumulant{\tau^2} / \cumulant{\tau}^3$\\ (if steady state is unique)} \\
		Large-deviation fct. & $I(j)$ [LFT of $c(z)$] & $\bar I(\tau)$ [LFT of $\bar c(s)$] & $I(j) = j\bar I(1/j)$ and $\bar I(\tau) = \tau I(1/\tau)$ \\ \addlinespace[-3pt]
		& \multicolumn{2}{l}{(LFT: Legendre-Fenchel transform)} & (derived in Ref.~\cite{GingrichPhysRevLett2017}) \\
		\bottomrule
	\end{tabular*}
	\caption{
		Summary of correspondences between full counting statistics and first-passage times in this paper.
		The second and third columns list quantities in the full counting statistics approach and analogous quantities in the first-passage time approach.
		The fourth column provides relations between them.
		In the main text, these relations are explained in more detail, including discussions of situations where they break down, and the generalization to two-way currents in Sec.~\ref{sec:bidi}.
	}
	\label{tab:uni:correspondence}
\end{table*}

Using Eq.~\eqref{eq:fptd}, the moment-generating function becomes 
\begin{equation}
\bar c_n(s) = \tr \bar\rho_n(s),
\end{equation}
where $\bar\rho_n(s)$ is the Laplace transform of $\bar\rho_n(t)$.
Because of the convolutions in Eqs.~\eqref{eq:trajectory} and \eqref{eq:barrho_n}, the first-passage propagators in Laplace space take the simple form
\begin{equation} \label{eq:convolution}
	\bar\rho_n(s) = \op F_n(s) \rho_0 = \op F_1(s)^n\rho_0,
\end{equation}
and it follows from Eqs.~\eqref{eq:n-resolved-prop} and \eqref{eq:fnt} that $\op F_1(s) = \op J \op T_s(0)$ with
\begin{equation} \label{eq:convolution2}
	\op T_s(0) = ( s - \op L_0 )^{-1} .
\end{equation}
Note that this identity for $\op T_s(0)$ holds only in the one-way current case.
The result \eqref{eq:convolution} is rooted in the dynamics being Markovian.
It intuitively states that reaching the count $n$ is the same as reaching the count of one $n$ times.
We have thus derived the expression
\begin{equation}
	\bar c_n(s) = \tr\bigl[ \op F_1(s)^n \rho_0 \bigr]
\end{equation}
for the moment-generating function.
Compared to the full counting statistics, the propagator $\op F_1(s)$ plays a role similar to that of the tilted generator \eqref{eq:tilted_gen}, and the Laplace parameter $s$ is the corresponding counting field.

To expand the analogy between the two approaches, we consider the jump-time averaged states~\cite{GneitingPhysRevA2021}
\begin{equation} \label{eq:jump-time-state}
	\bar\rho_n = \int_0^\infty \dd t\, \bar\rho_n(t) = \op F_n(s=0) \rho_0.
\end{equation}
These states describe the system immediately after the $n$-th jump, while disregarding the information about the times when the jumps happened.
The operator $\op F_1 = \op F_1(s=0)$ describes the evolution of the jump-time averaged states from one jump count to the next as
\begin{equation} \label{eq:jump-time-state-evo}
	\bar\rho_{n+1} = \op F_1 \bar\rho_n.
\end{equation}
That is analogous to how the generator $\op L = \op L(z=1)$ generates the time evolution of the wall-time averaged states $\rho_t$.

If $\op F_1(s)$ has a unique eigenvalue with maximum absolute value (ignoring multiplicities), the moment-generating function grows exponentially for large $n$, and the scaled generating function is then given by this particular eigenvalue as
\begin{equation} \label{eq:fps-sgf}
	\bar c(s) = \lim_{n \to \infty} \sqrt[n]{\bar c_n(s)} = \lambda_{\mathrm{max}}^{\mathrm{abs}}[ \op F_1(s) ].
\end{equation}
In analogy to our assumption that the wall-time averaged state cannot exhibit coherent oscillations at long times, we also assume that there are no oscillations in the jump-time evolution for large $n$.
As shown in App.~\ref{app:oscill}, those two assumptions are not necessarily equivalent.
Mathematically, the additional assumption requires that the quantum channel $\op F_1$ has no eigenvalues on the unit circle other than the eigenvalue one.
In Sec.~\ref{subsec:cumulant_relations}, we will see that $\bar c(s)$ is a smooth function in a neighborhood of $s=0$ if and only if $c(z)$ is smooth around $z=1$. In that case, the scaled cumulants of the first-passage time are
\begin{equation} \label{eq:fpt_cumulants}
	\cumulant{\tau^k}=\lim_{n \to \infty} \frac 1 n\, \cumulant{t^k}_n = (-\partial_s)^k \ln\bigl[ \bar c(s) \bigr] \bigr|_{s=0}.
\end{equation}

\subsection{Finite-Time Correspondence} \label{subsec:main_thm}

We are now ready to discuss the correspondence between the two approaches.
First, we introduce a relationship between the operators $\op L(z)$ and $\op F_1(s)$, which are the dynamical generators in the full counting statistics and the first-passage time approach, respectively.
In the subsequent sections, we then discuss its implications.
The relationship generalizes the asymptotic correspondence between the scaled generating functions from Ref.~\cite{BudiniJStatMech2014} by extending it to arbitrary times and thereby providing a complete picture of the ensemble relation.
The various types of correspondence discussed here and in the following are summarized in Table~\ref{tab:uni:correspondence}.

The key equation is the identity
\begin{equation} \label{eq:budini}
	\bigl[ \op F_1(s) - z \bigr] \op T_s(0)^{-1} = z \bigl[ \op L(z) - s \bigr],
\end{equation}
which can be verified by using that $\op F_1(s) = \op J \op T_s(0)$, which follows from the definition \eqref{eq:fnt}, and inserting the expressions \eqref{eq:tilted_gen} and \eqref{eq:convolution2} for $\op L(z)$ and $\op T_s(0)$.
The identity does not apply for $z=0$ and for values of $s$ which are eigenvalues of $\op L_0$, since $\op L(z)$ and $\op F_1(s)$ are undefined at those values.
We will explore the consequences of this fact later in this section.

The following four statements are an immediate consequence of the key equation above.
They characterize the finite-time correspondence between the generators.
\begin{enumerate}[(1), leftmargin=*]
\item If $\op F_1(s)$ has an eigenvalue $z$, then either $z$ is zero or $\op L(z)$ has the eigenvalue $s$.
\item If $z \neq 0$ is an eigenvalue of $\op F_1(s)$ with right and left eigenstates $\bar\rho$ and $\bar\varphi$, the corresponding right and left eigenstates of $\op L(z)$ are $\rho = \op T_s(0) \bar\rho$ and $\varphi = \bar\varphi$.
	Here, we view left eigenstates as linear functionals on the state space.
\item If $\op L(z)$ has an eigenvalue $s$, then either $s$ is an eigenvalue of $\op L_0$ or $\op F_1(s)$ has the eigenvalue $z$.
\item If $s$ is not an eigenvalue of $\op L_0$ but an eigenvalue of $\op L(z)$ with right and left eigenstates $\rho$ and $\varphi$, the corresponding right and left eigenstates of $\op F_1(s)$ are $\bar\rho = \op T_s(0)^{-1} \rho$ and $\bar\varphi = \varphi$.
\end{enumerate}
We note that Eq.~\eqref{eq:budini} was also formulated in Ref.~\cite{BudiniJStatMech2014}, but that these consequences were not explored there.

Our four statements show that the spectral plots of the generators $\op L(z)$ and $\op F_1(s)$ are mirror images of each other, with some well-defined exceptions.
By spectral plot, we mean a plot of all eigenvalues of the operator as a function of the respective counting field.
Examples of such plots can be found in Figs.~\ref{fig:ex1} and \ref{fig:ex2} and are discussed in subsequent sections.
The eigenstates associated with corresponding points in these spectral plots are also related, with the left eigenstates even being the same.

The two possible exceptions to the symmetry of the spectral plots are zero eigenvalues of $\op F_1(s)$ and shared eigenvalues of $\op L(z)$ and $\op L_0$.
To understand when the first type of exception happens, we first recognize that $\varphi$ is a left eigenstate of $\op F_1(s)$ with eigenvalue zero, such that $\varphi\, \op F_1(s) = 0$, if and only if $\varphi \op J = 0$.
Therefore, $\op F_1(s)$ has a zero eigenvalue with multiplicity equal to the rank deficiency of $\op J$.
The rank deficiency of an operator is the number of its zero eigenvalues.
Physically, a rank deficient jump operator means that the states that are possible jump destinations do not span the entire state space; hence, these zero eigenvalues indicate that jump events can be accompanied by information gain about the system state.

The second type of exception, where $\op L(z)$ and $\op L_0$ share an eigenvalue, does not have a simple characterization like the first type.
However, dark states, meaning states  with $\op L_0 \rho = 0$, lead to constant zero eigenvalues of $\op L(z)$, and if there are no dark states, then $\op L(z)$ cannot have any constant zero eigenvalue. These statements follow from a more elaborate discussion in Appendix~\ref{app:dark_states}.

We have now concluded our discussion of the symmetry between the eigendecompositions of the generators $\op L(z)$ and $\op F_1(s)$, and the possible exceptions.
Outside of these exceptions, our results imply that either generator is fully determined by the other, since operators are fully determined by their eigendecompositions.
In the remainder of this section, we investigate how the exceptions change this picture.

Our discussion of the first type of exception shows that the eigenvalues and eigenstates of $\op F_1(s)$ are fully determined by those of $\op L(z)$ and $\op J$.
Since $\op J$ can be asymptotically obtained from $\op L(z)$ in the limit $z \to 0$, we conclude that $\op F_1(s)$ is still fully determined by $\op L(z)$ irrespective of the exceptions.
Concretely, it can be written in the diagonalized form
\begin{equation} \label{eq:uni:F_explicit}
	\op F_1(s) = \op X(s)^{-1} \op \Lambda(s)\, \op X(s) ,
\end{equation}
where $\Lambda(s)$ is a diagonal matrix containing the eigenvalues of $\op F_1(s)$, and $\op X(s)$ a basis transformation matrix, whose rows are the left eigenstates of $\op F_1(s)$.
Specifically, $\op\Lambda(s)$ contains a zero eigenvalue with multiplicity equal to the rank deficiency of $\op J$, and the other eigenvalues are those values of $z$ where $\op L(z)$ has the eigenvalue $s$.
The left eigenstates of $\op F_1(s)$, which are contained in $\op X(s)$, are the left eigenstates of $\op J$ and of $\op L(z)$ corresponding to the respective eigenvalues.
Equation~\eqref{eq:uni:F_explicit} provides an alternative way of computing $\op F_1(s)$.
Here, in the one-way current case, it is generally easier to apply Eqs.~\eqref{eq:fnt} and \eqref{eq:convolution2} instead.
However, we will see later that Eq.~\eqref{eq:uni:F_explicit} generalizes to the two-way current case, whereas Eq.~\eqref{eq:convolution2} does not.

In the generic case, also $\op L(z)$ is fully determined by $\op F_1(s)$.
However, interestingly, it is possible in some cases that $\op F_1(s)$ does not contain all the information required to construct $\op L(z)$.
We illustrate this issue in Appendix~\ref{app:quick_example} with a simple example, where the dynamics of the system depends on a rate $\alpha$, but the first-passage propagator is independent of this rate.
As discussed in more detail in Appendix~\ref{app:quick_example}, this situation can only happen if there is a part of the dynamics that influences neither the first-passage time distributions nor the conditioned jump-time averaged states.
Such a part of the dynamics cannot influence the first-passage propagator, which only knows about the first-passage times and the jump-time states.
In contrast, the tilted generator contains the full information about the system by definition.
Formally, our correspondence shows that this situation occurs if and only if $\op L(z)$ has a counting-field independent eigenvalue, since the corresponding line in the spectral plot of $\op F_1(s)$ would be vertical, which is impossible.
This counting-field independent eigenvalue must then also be an eigenvalue of $\op L_0$.

\subsection{Steady States and Observables} \label{subsec:steady_states}

Next, we consider the steady states in the thermodynamic limit of long times and large jump counts.
To ensure that large counts are reached, we assume that there are no dark states.
Furthermore, we still assume that the system does not exhibit oscillations at long times in either the wall-time or the jump-time picture.
Both the regular system state $\rho_t$ and the jump-time averaged state $\bar\rho_n$ then converge to steady states for large $t$ and $n$.
Those may be nonequilibrium states, they may be different from each other, and they may depend on the initial conditions.
We refer to states $\rho$ with $\op L\rho = 0$ as wall-time steady states, and to states $\bar\rho$ with $\op F_1 \bar\rho = \bar\rho$ as jump-time steady states.
Generally, the set of jump-time steady states depends on the choice of the jump operator in Eq.~\eqref{eq:jump_op}, while the set of wall-time steady states does not.

Because the correspondence above connects the eigenstates of the operators $\op L(z)$ and $\op F_1(s)$, the steady states are also related, since they indeed are eigenstates.
This relation is one-to-one: for each jump-time steady state $\bar\rho$, we obtain a wall-time steady state $\rho$ as
\begin{equation} \label{eq:steady_states2}
	\rho \propto \op T_{s=0}(0) \bar\rho = -\op L_0^{-1} \bar\rho = \int_0^\infty \dd t\, \ee^{\op L_0 t} \bar\rho ,
\end{equation}
where proportional means equality up to normalization.
The fixed-time ensemble $\rho$ can thus be obtained from the ensemble $\bar\rho$, which consists of states at different times, by applying the no-jump propagators $\exp(\op L_0 t)$.
This operation can be interpreted as propagating the different parts of the ensemble to the same time.

Conversely, for a wall-time steady state~$\rho$, we obtain the corresponding jump-time steady state as
\begin{equation} \label{eq:steady_states}
	\bar\rho \propto \op T_{s=0}(0)^{-1} \rho = -\op L_0 \rho = \op J \rho .
\end{equation}
In the last equality, we used $\op L\rho = 0$.
The resulting expression, which was previously found in Ref.~\cite{LandiArXiv230507957Quant-Ph2023}, is the (unnormalized) state right after a quantum jump, given that the system was in the state $\rho$ before the jump.
This result can be given an operational interpretation as follows.
The wall-time steady state $\rho$ can be prepared by waiting for a long time without observing the system.
If, after starting the observation, a jump event happens immediately, the state will be in the jump-time steady state $\bar\rho$ after.
The state $\bar\rho$ could thus be prepared in a post-selection process, where one first prepares $\rho$ by waiting, then begins the observation of jump events, and finally discards the experiment if a jump event does not happen immediately within a sufficiently small time $\dd t$.

For a given initial condition $\rho_0$, the wall-time and jump-time evolutions lead to the steady states $\rho_\infty$ and~$\bar\rho_\infty$.
Even though the operators $\op T_{s=0}(0)$ and its inverse are maps between wall-time and jump-time steady states, they generally do not map $\rho_\infty$ and $\bar\rho_\infty$ onto each other.
That is, the jump-time steady state $\bar\rho_\infty$ is generally not proportional to $\op J \rho_\infty$.
However, if the steady states are unique, we do have $\bar\rho_\infty \propto \op J \rho_\infty$ for any initial state.

To explore the general relationship between $\rho_\infty$ and $\bar\rho_\infty$, it is helpful to investigate the left eigenstates corresponding to the eigenvalues zero of $\op L$ or one of $\op F_1$.
Following the finite-time correspondence between the generators, the set of left eigenstates is the same for the two.
A given basis $\rho_i$ spanning the wall-time steady states uniquely determines a basis $\varphi_i$ of the left eigenstates through the orthonormality condition $\varphi_j \rho_i = \delta_{ij}$.
The corresponding basis $\bar\rho_i \propto \op J \rho_i$ fixes another basis $\bar\varphi_i$ of the left eigenstates through the relation $\bar\varphi_j \bar\rho_i = \delta_{ij}$. Generally, the two bases are not the same.
However, as we show in Appendix~\ref{app:steadystates}, there exists a particular basis $\rho_i$ such that
\begin{equation} \label{eq:left_eigenstates}
	\bar\varphi_i = \varphi_i.
\end{equation}
We now fix the basis $\rho_i$ to be that particular basis.

Given an initial state $\rho_0$, the dynamically obtained wall-time and jump-time steady states are
\begin{align}
	\rho_\infty &= \sum_i \lambda_i \rho_i
\shortintertext{and}
	\bar\rho_\infty &= \sum_i \lambda_i \bar\rho_i,
\end{align}
where $\lambda_i = \varphi_i \rho_0 = \bar\varphi_i \rho_0$.
We now see that $\bar\rho_\infty \propto \op J \rho_\infty$ holds only if $\tr[ \op J \rho_i ]$ has the same value for all $i$ with $\lambda_i \neq 0$.
If these values are all different, we have $\bar\rho_\infty \propto \op J \rho_\infty$ only if $\rho_\infty$ is among the particular basis states.

Before concluding this section, we return to systems with unique steady states. As mentioned previously, $\op T_{s=0}(0)^{-1} = -\op L_0$ necessarily maps $\rho_\infty$ to $\bar\rho_\infty$ in this situation.
It is then possible to introduce a correspondence between observables in the two approaches.
If $A$ is an observable measured in the wall-time steady state, its expectation value is $\expval{A}_{\text{wt}} = \tr[A \rho_\infty]$.
Measuring the same observable in the jump-time steady state generally yields a different result.
However, the observable
\begin{equation} \label{eq:obs_correspondence}
	\bar A = -j_{\text{wt}}\, (\op L_0^\dagger)^{-1} A,
\end{equation}
measured in the jump-time steady state, is equivalent to $A$ in the sense that it has the same expectation value
\begin{equation}
	\expval{\bar A}_{\text{jt}} = \tr[\bar A \bar\rho_\infty] = \tr[A \rho_\infty] = \expval{A}_{\text{wt}} .
\end{equation}
Above, $j_{\text{wt}} = \tr[\op J \rho_\infty]$ is the average current in the wall-time steady state, and $\op L_0^\dagger$ is the adjoint operator.

This result shows that the steady states in the two approaches carry the same information, in the sense that for any observable measured in either one approach, we can in principle construct an equivalent observable in the other approach.
We note, however, that this equivalent observable may not have an immediate physical interpretation. Furthermore, finite-time observables, such as the topological current considered in Ref.~\cite{GneitingPhysRevRes2022}, may not have an equivalent.

Above, we restricted ourselves to systems with unique steady states, since $\op T_{s=0}(0)$ is then a canonical choice for a map that maps $\bar\rho_\infty$ to $\rho_\infty$, is bijective and preserves Hermiticity.
Because of these properties, the prescription \eqref{eq:obs_correspondence} gives rise to a one-to-one mapping between the observables $A$ and $\bar A$.
In systems with multiple steady states, any map with the same properties also induces a correspondence between observables, but there is no canonical choice.
We will see an example of such a map in an application in Sec.~\ref{sec:exampleSSH}.

\subsection{Scaled Cumulants and Metastability} \label{subsec:cumulant_relations}

In this section, we investigate the relationship between the scaled cumulants of the full counting statistics and of the first-passage time.
Under the assumptions made in the previous section, the eigenvalue zero of $\op L$ and the eigenvalue one of $\op F_1$ are maximal with respect to the real part and to the absolute value, respectively.
The scaled generating functions are then defined in neighborhoods of $z=1$ and $s=0$ where there are no eigenvalue crossings.
They satisfy $c(1) = 0$ and $\bar c(0) = 1$ and are inverse functions of each other on that neighborhood:
\begin{align}
	c(\bar c(s)) &= s \label{eq:budini_gf1}
\shortintertext{and}
	\bar c(c(z)) &= z . \label{eq:budini_gf2}
\end{align}
This relationship was first pointed out in Ref.~\cite{BudiniJStatMech2014}.

The meaning of this formal relation can be made concrete in terms of the large-deviation functions.
The large-deviation functions describe the asymptotic probabilities of measuring an output current $j=n/t$ after a long time,
\begin{align}
	I(j) &= \lim_{t \to \infty} \frac{\ln P_t(jt)}{t} , \label{eq:ldf_definition_j}
\intertext{or measuring an average first-passage time $\tau=t/n$,}
	\bar I(\tau) &= \lim_{n \to \infty} \frac{\ln \bar P_n(\tau n)}{n}, \label{eq:ldf_definition_tau}
\end{align}
after many jumps.
According to the Gärtner-Ellis theorem, the large-deviation functions are related to the scaled generating functions through Legendre-Fenchel transformations~\cite{TouchettePhysRep2009}.
The inverse function relationship then translates to the identities~\cite{GingrichPhysRevLett2017}
\begin{align}
	I(j) &= j \bar I(1 / j) \label{eq:ldf_identity2}
\shortintertext{and}
	\bar I(\tau) &= \tau I(1 / \tau) \label{eq:ldf_identity}
\end{align}
for the large-deviation functions.
One can thus interpret Eqs.~\eqref{eq:budini_gf1} and \eqref{eq:budini_gf2} as connecting the asymptotic probabilities of measuring the current $j$ and the average first-passage time $\tau = 1/j$.

In the case of unique steady states, the scaled cumulants of the two approaches are encoded in derivatives of the scaled generating functions.
Thus, one can take derivatives of the inverse function relationship to find relations between the two sets of scaled cumulants.
The first two such relations are
\begin{align}
	\cumulant{j} &= 1/\cumulant{\tau} \label{eq:cumulant_relations1}
\shortintertext{and}
	\cumulant{j^2} &= \cumulant{\tau^2} / \cumulant{\tau}^3 , \label{eq:cumulant_relations2}
\end{align}
and similar expressions can be derived for the higher cumulants.
The left-hand sides of these equations contain scaled cumulants of the full counting statistics, and the right-hand side those of the first-passage time statistics.

Any constraints on the cumulants in one approach can thus be translated into constraints in the other approach.
This technique was applied to the kinetic uncertainty relation in Ref.~\cite{GarrahanPhysRevE2017}, and the result was extended to finite times in Ref.~\cite{HiuraPhysRevE2021}.
Similarly, Ref.~\cite{GingrichPhysRevLett2017} applied the technique to translate the thermodynamic uncertainty relation~\cite{BaratoPhysRevLett2015, GingrichPhysRevLett2016, SeifertPhysicaA2018, HorowitzNatPhys2020} to the first-passage time picture.
This relation is nontrivial only for two-way currents; however, we will see that the cumulant relations \eqref{eq:cumulant_relations1} and \eqref{eq:cumulant_relations2} still hold in that case.
Other kinetic and thermodynamic uncertainty relations in the first-passage time picture have been formulated in Refs.~\cite{PalPhysRevRes2021, HasegawaPhysRevE2022, Bakewell-SmithJStatPhys2025, RaghuNewJPhys2025, MacieszczakArXiv240709839Cond-Matstat-Mech2024}.

\begin{figure*}[t]
	\includegraphics[scale=1,valign=t]{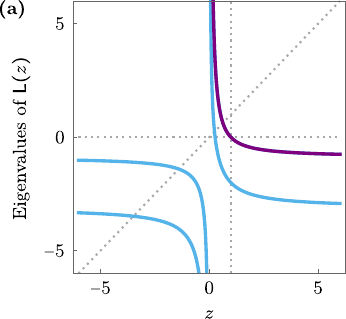}%
	\hfill%
	\includegraphics[scale=1,valign=t]{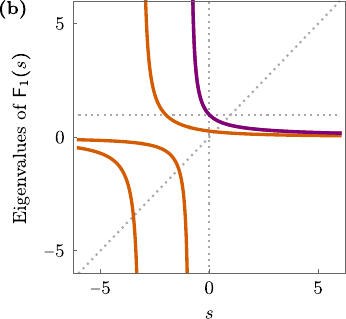}%
	\hfill%
	\includegraphics[scale=1,valign=t]{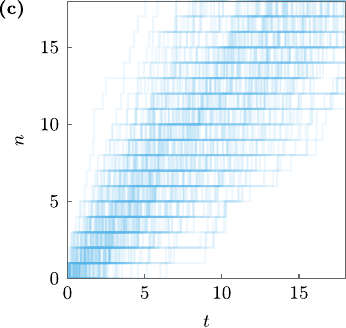}%
	\caption[]{
        Two-state emitter.
        (a,b) Spectral plots of the generators $\op L(z)$ and $\op F_1(s)$.
        The points $(z{=}1,s{=}0)$, where derivatives are taken to find the cumulants at long times, are indicated by dotted lines.
        The largest eigenvalues around these points (highlighted in purple) are the scaled generating functions $c(z)$ and $\bar c(s)$.
        The inverse function relationship derived in Ref.~\cite{BudiniJStatMech2014} can be seen in the mirror symmetry between them.
        Our generalized, finite-time ensemble correspondence shows that not only $c(z)$ and $\bar c(s)$, but the entire spectral plots (a) and (b) are mirror images of each other.
		The mirror symmetry is indicated by diagonal dotted lines.
        (c) Number of emitted photons as a function of time for 150 randomly generated trajectories.
        The result shows that in this example involving only one-way currents, the counted number can only increase.
        In all panels, we use arbitrary units with $\gamma_1 = 1/2$, $\gamma_2 = 2$, and $\Gamma = 1$.
	}
	\label{fig:ex1}
\end{figure*}

We now focus on the situation where the steady states are not unique, and they then depend on the initial condition. The eigenvalue zero of $\op L$ and the eigenvalue one of $\op F_1$ correspondingly have multiplicities larger than one. In that case, the scaled cumulant generating functions generally do not depend smoothly on the counting fields at $z=1$ and $s=0$. Also, the cumulant relations \eqref{eq:cumulant_relations1} and  \eqref{eq:cumulant_relations2} can be violated however intuitive they may appear.

As an example, we consider a system with two steady states, and we assume that we can diagonalize the operators $\op L(z)$ and $\op F_1(s)$.
We denote the two dominating normalized eigenstates of $\op L(z)$ as $\rho_i(z)$  and those of $\op F_1(s)$ as $\bar\rho_i(s)$ with $i \in \{1,2\}$.
They satisfy the eigenvalue equations
\begin{align}
	\op L(z) \rho_i(z) &= \ell_i(z) \rho_i(z)
\shortintertext{and}
	\op F_1(s) \bar\rho_i(s) &= f_i(s) \bar \rho_i(s).
\end{align}
Here, the two eigenvalues $\ell_i(z)$ and $f_i(s)$ are smooth functions of the counting fields, and they satisfy $\ell_i(1) = 0$ and $f_i(0) = 1$.
They are also inverse functions of each other, and therefore we have $f_i'(0) = 1 / \ell_i'(1)$, where the primes denote derivatives.
The states $\rho_i(1)$ are wall-time steady states, and $\bar\rho_i(0)$ jump-time steady states.

We now expand the initial state of the system both in the eigenstates of $\op L(z)$ and in those of $\op F_1(s)$ as
\begin{align}
	\rho_0 &= \lambda_1(z) \rho_1(z) + \lambda_2(z) \rho_2(z) + \dots \nonumber\\
	&= \bar\lambda_1(s) \bar\rho_1(s) + \bar\lambda_2(s) \bar\rho_2(s) + \dots ,
\end{align}
where $\lambda_i(z)$ and $\bar\lambda_i(s)$ are the expansion coefficients, and the dots stand for the parts of the expansions involving all other eigenstates.
Our discussion in the previous section shows that $\lambda_i(1) = \bar\lambda_i(0)$, and we set $\lambda_i = \lambda_i(1) = \bar\lambda_i(0)$. We also note that $\lambda_1 + \lambda_2 = 1$.

We can then find the two generating functions.
To leading order in $t$ and $n$, contributions from all other eigenvalues can be disregarded, and we obtain
\begin{align}
	c_t(z) &= \ln\bigl[ \lambda_1(z) \ee^{\ell_1(z) t} + \lambda_2(z) \ee^{\ell_2(z) t} \bigr]
\shortintertext{and}
	\bar c_n(s) &= \bar\lambda_1(s)\, f_1(s)^n + \bar\lambda_2(s)\, f_2(s)^n .
\end{align}
For the first moments, we then find
\begin{align}
	\expval{n}_t &= -\partial_z c_t(z) \bigr|_{z=1} = - \bigl( \lambda_1 \ell_1'(1) + \lambda_2 \ell_2'(1) \bigr) t \label{eq:metastable_moments1}
\intertext{and}
	\expval{t}_n &= -\partial_s \ln\bigl[ \bar c_n(s) \bigr] \bigr|_{s=0} = - \bigl( \lambda_1 \ell_1'(1)^{-1} + \lambda_2 \ell_2'(1)^{-1} \bigr)n . \label{eq:metastable_moments2}
\end{align}
Assuming that $\ell_1'(1) \neq \ell_2'(1)$ and that both $\lambda_i$ are non-zero, the cumulant relation \eqref{eq:cumulant_relations1} is violated.
In this case, all higher cumulants grow faster than linearly in $t$ or $n$, making the scaled cumulants diverge. This divergence has also been found for multistable systems~\cite{JordanPhysRevLett2004, SchallerPhysRevB2010, KarzigPhysRevB2010, DegerPhysRevE2018}.

\begin{figure*}[t]
	\centering
    \includegraphics[scale=1]{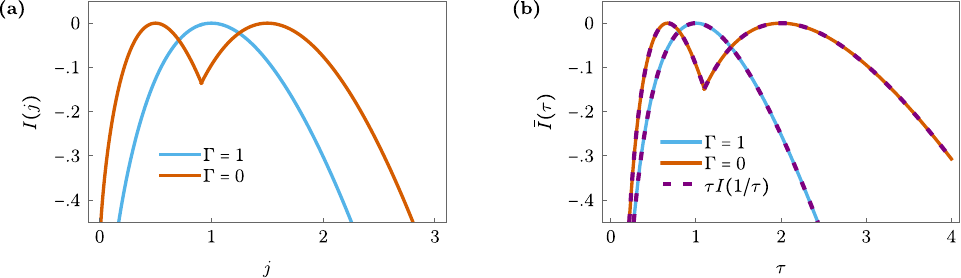}
	\caption[]{
		Large-deviation functions for the two-state emitter.
        \begin{enumerate*}[(a)]
        \item Large-deviation function for the full counting statistics.
        \item Large-deviation function for the first-passage time.
        \end{enumerate*}
        The results in this figure illustrate the relationship \eqref{eq:ldf_identity} between the large-deviation functions \cite{GingrichPhysRevLett2017}.
        By transforming the large-deviation function in panel (a) according to Eq.~\eqref{eq:ldf_identity}, we obtain the dashed lines in panel (b), which match the large-deviation function for the first-passage time.
        The large-deviation functions are shown for two values of $\Gamma$, featuring a unique steady state ($\Gamma = 1$) and multiple steady states ($\Gamma = 0$), highlighting that this relationship continues to hold even without a unique steady state.
		The remaining parameters are the same as in Fig.~\ref{fig:ex1}.
	}
	\label{fig:ex1:ldfs}
\end{figure*}

In the absence of fundamental symmetries, physical systems typically do not have multiple steady states. However, in reality, experimental observation times are also not infinite. Whether a system effectively exhibits multiple steady states depends on whether there are metastable states with lifetimes longer than the observation time. In a system with metastability, one may find a regime, where the first moments of the full counting and the first-passage statistics are given by Eqs.~\eqref{eq:metastable_moments1} and \eqref{eq:metastable_moments2} for observation times that are shorter than the lifetimes of the metastable states, but longer than all other time scales.
For observation times that are even longer than the metastable lifetimes, the cumulant relations in Eqs.~\eqref{eq:cumulant_relations1} and Eqs.~\eqref{eq:cumulant_relations2} hold.
In between, there is a crossover between the two behaviors. We provide an example of this phenomenon in the following section.

Therefore, violations of the cumulant relations indicate the presence of metastable states. In practice, metastable states can therefore be detected by measuring and comparing both sides of these relations.
Any observed violations must stem from metastable states; however, not observing a violation does not exclude metastable states.

\subsection{Application: Two-State Emitter} \label{subsec:exampleTWE}

As an application of our formalism, we now consider an emitter with two internal states that we label by ``1'' and ``2'', as illustrated in Fig.~\ref{fig:examples}(a).
The emitter switches stochastically between the states at the rate $\Gamma$ and emits photons at the rates $\gamma_1$ or $\gamma_2$, depending on the state of the system. We can describe the system using a classical rate equation, where the photon emissions are included as transitions from each state to itself at the rates $\gamma_{ii} = \gamma_i$.
We are thus able to write down the operators
\begin{equation}
	\op L_0 = \begin{pmatrix} -\gamma_1 - \Gamma & \Gamma \\ \Gamma & -\gamma_2 - \Gamma \end{pmatrix}
\end{equation}
and
\begin{equation}
	\op J = \begin{pmatrix} \gamma_1 & 0 \\ 0 & \gamma_2 \end{pmatrix}.
\end{equation}

With this example, we first illustrate the correspondence between generators discussed in Sec.~\ref{subsec:main_thm} by considering the spectral plots in Figs.~\ref{fig:ex1}(a) and \ref{fig:ex1}(b).
Here we see that each plot is the mirror image of the other, and it can be shown that for each pair of points related by the symmetry, the corresponding left eigenvectors are the same, and the right eigenvectors are related by~$\op T_s(0)$.

In both panels, we highlight the maximum eigenvalues in the neighborhoods of $z=1$ or $s=0$ where they are inverse functions.
However, setting $c(z) = \lambda_{\mathrm{max}}^{\mathrm{re}}[ \op L(z) ]$ and $\bar c(s) = \lambda_{\mathrm{max}}^{\mathrm{abs}}[ \op F_1(s) ]$ for all $z$ and $s$ would not lead to globally inverse functions.
The relationship between the maximum eigenvalues, which are the scaled generating functions, is equivalent to Eqs.~\eqref{eq:ldf_identity2} and \eqref{eq:ldf_identity} for the large-deviation functions, which is illustrated in Fig.~\ref{fig:ex1:ldfs}.
In contrast to the cumulant relations in Eqs.~\eqref{eq:cumulant_relations1} and \eqref{eq:cumulant_relations2}, which we investigate below, this identity remains valid even at zero switching rate, $\Gamma=0$, where the steady state is not unique.

On timescales that are longer than the inverse switching rate, the trajectories of the system are likely to sample both system states.
For each trajectory, the average rate of emissions is thus approximately $(\gamma_1 + \gamma_2) / 2$.
This argument motivates why the  relation in Eq.~\eqref{eq:cumulant_relations1} should hold. Indeed, a direct calculation confirms that
\begin{align}
	\frac{\expval{n}_t}{t} &= -\frac 1 t \partial_z c_t(z) \bigr|_{z=1} = \frac{\gamma_1 + \gamma_2}{2} \label{eq:ex1:regime1}
\intertext{and}
	\frac{\expval{t}_n}{n} &= -\frac 1 n \partial_s \ln \bar c_n(s)\bigr|_{s=0}= \frac{2}{\gamma_1 + \gamma_2} \label{eq:ex1:regime1_2}
\end{align}
to leading order in large $\Gamma t$ or $n$, and that the analogous relations for the higher cumulants hold as well.

On timescales that are shorter than the inverse switching rate, and for jump counts smaller than $\gamma_1/\Gamma$ and $\gamma_2 / \Gamma$, the trajectories are likely to remain in one of the two states.
Depending on the initial condition, there is thus a fraction $\lambda_1$ of trajectories with the average emission rate~$\gamma_1$, and a fraction $\lambda_2 = 1 - \lambda_1$ of trajectories, where the rate is $\gamma_2$.
Averaging the jump count and the first-passage time over all trajectories yields the moments 
\begin{align}
	\frac{\expval{n}_t}{t} &= \lambda_1 \gamma_1 + \lambda_2 \gamma_2 \label{eq:ex1:regime2}
\shortintertext{and}
	\frac{\expval{t}_n}{n} &= \lambda_1 \gamma_1^{-1} + \lambda_2 \gamma_2^{-1} \label{eq:ex1:regime2_2}
\end{align}
which is also expected according to Eqs.~\eqref{eq:metastable_moments1} and \eqref{eq:metastable_moments2}.

\begin{figure}[t]
    \centering
	\includegraphics[scale=1]{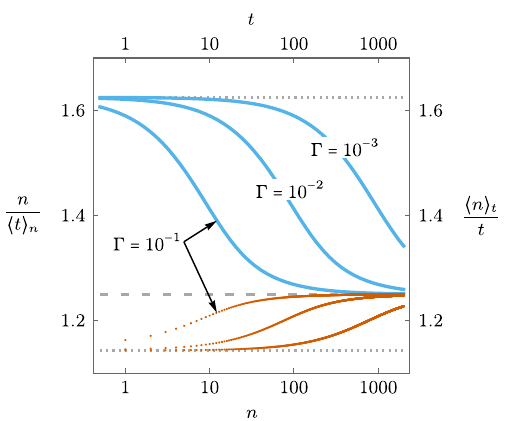}
	\caption{
        Two-state emitter.
        We show the average current, $\expval{n}_t / t$, as a function of the observation time $t$ (blue solid lines) together with the average first-passage time per jump, $\expval{t}_n / n$, as a function of the discrete target count $n$ (orange points).
        These results show that the emitter becomes metastable as the switching rate $\Gamma$ becomes smaller.
        For small switching rates, we find a crossover between small $t$ and $n$, where the emitter effectively has multiple steady states and the results can be described approximately by Eqs.~\eqref{eq:ex1:regime2} and \eqref{eq:ex1:regime2_2} (dotted gray lines), and large $t$ and $n$, where the emitter reaches the true steady states and the results match Eqs.~\eqref{eq:ex1:regime1} and \eqref{eq:ex1:regime1_2}.
		We have used the same parameters as in Fig.~\ref{fig:ex1} with the initial state $\rho_0 = (1/4, 3/4)$.
	}
	\label{fig:ex1_metastability}
\end{figure}

Figure~\ref{fig:ex1_metastability} illustrates the crossover between the two regimes.
For small switching rates, the system is metastable, and Eqs.~\eqref{eq:ex1:regime2} and \eqref{eq:ex1:regime2_2} provide an accurate description over a wide range of observation times $t$ and target counts $n$.
One can thus detect the metastability from the violation of the cumulant relation in Eq.~\eqref{eq:cumulant_relations1}.

\subsection{Application: SSH Model} \label{sec:exampleSSH}

\begin{figure*}[t]
	\includegraphics[scale=1,valign=t]{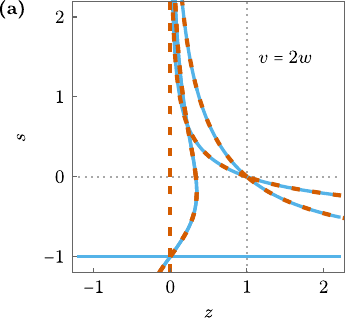}%
	\hfill%
	\includegraphics[scale=1,valign=t]{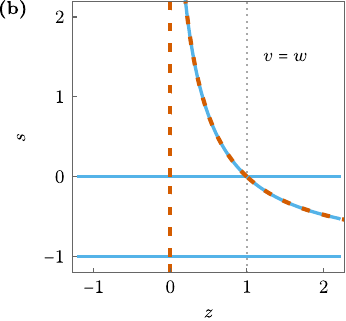}%
	\hfill%
	\includegraphics[scale=1,valign=t]{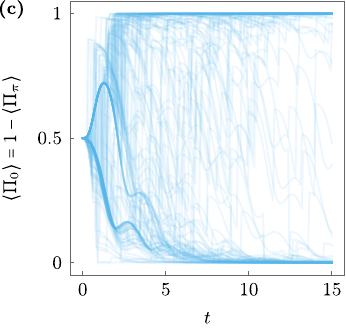}%
	\caption{
		SSH model.
        (a,b) Spectral plots for different ratios of $v/w$.
        The two panels show the eigenvalues of $\op L(z)$ (solid blue lines) and the mirrored eigenvalues of $\op F_1(s)$ (dashed orange lines).
        There are $16$ eigenvalues, but some are imaginary, outside the visible area, or have multiplicity larger than one.
        These spectral plots illustrate exceptions from the mirror symmetry, where one of the solid blue lines does not have a matching dashed orange line, or vice versa.
        Exceptions occur in panel (a) at $z=0$ and $s=-1$, and in panel (b) additionally at $s=0$, indicating the dark state present at $v=w$.
        (c)~To investigate the decoherence of a superposition state, we randomly generated $150$ trajectories, starting from the initial state $\frac{1}{\sqrt 2} \bigl( \ket{0} + \ket{\pi} \bigr) \otimes \ket B$.
		We show the expectation value $\expval{\Pi_0}$ along each trajectory, where $\Pi_k = \ketbra k k \otimes \mathbbm 1$.
		The result shows that this expectation value approaches either $0$ or $1$ on each individual trajectory, demonstrating the localization of the trajectory in the $k=\pi$ or the $k=0$ subspace, respectively.
		In all panels, we have used arbitrary units with $v=1$, $\gamma = 2$, and (a) $w = 1/2$ (b) $w = 1$ (c) $w = 1/2$.
	}
	\label{fig:ex2}
\end{figure*}

As our next application, we consider the Su--Schrieffer--Heeger (SSH) model, which describes a one-dimensional lattice with two sites per unit cell, connected by hopping amplitudes as indicated in Fig.~\ref{fig:examples}(b).
We also include a collective dissipation process as described below.
For the sake of simplicity, we consider only two unit cells and impose periodic boundary conditions. The restriction to two unit cells makes the following calculations tractable without significantly affecting the underlying physics.

States with different momenta decouple in the SSH model~\cite{Asboth2016}.
With two unit cells, the momentum only takes the values $k = 0, \pi$ (with the lattice constant set to one).
In terms of the momentum eigenstates $\ket k$, the Hamiltonian can then be written as 
\begin{align}
	H &= \sum_k \left(\ketbra k k \otimes H_k\right) ,
\shortintertext{where}
	H_k &= \bigl[ v + w \cos(k) \bigr]\sigma_x - w \sin(k) \sigma_y
\end{align}
acts on the intracell basis $\{ \ket A, \ket B \}$.
Here, the tunneling amplitudes are denoted by $v$ and $w$, as shown in Fig.~\ref{fig:examples}(b), and $\sigma_{x,y}$ are the Pauli matrices.
The decay channel is described by the rate $\gamma$ and the Lindblad operator
\begin{equation}
	L = \sum_k \left(\ketbra k k \otimes \sigma_-\right) = \mathbbm 1_k \otimes \sigma_-,
\end{equation}
where $\sigma_- = \ketbra A B$.
Hence, the free generator reads
\begin{equation}
	\op L_0 \rho = -\ii \comm{H}{\rho} + \frac \gamma 2 \acomm[\big]{L^\dagger L}{\rho} ,
\end{equation}
while the jump operator takes the form
\begin{equation}
	\op J \rho = \gamma\, L \rho L^\dagger.
\end{equation}

In Figs.~\ref{fig:ex2}(a) and~\ref{fig:ex2}(b), we again show spectral plots of the operators $\op L(z)$ and $\op F_1(s)$.
Due to the decoupling of the two momentum subspaces, the system has two steady states, which follows from the eigenvalue crossings at $z=1$ and $s=0$ in Fig.~\ref{fig:ex2}(a).
The maximum eigenvalues as functions of $z$ and $s$ are not differentiable at that point, and the cumulant relations are not fulfilled.
For $v=w$, one of the steady states becomes a dark state.
Hence, there is no mirror symmetry in Fig.~\ref{fig:ex2}(b) at the point where $z=1$ and $s=0$.
In that case, the generating functions are not related by inversion.
The dark state appearing for $v=w$ is a signature of the topological phase transition in the infinite-lattice SSH model~\cite{Asboth2016, GneitingPhysRevRes2022}.

For the two-state emitter, the violation of the cumulant relations occurs because all trajectories remain in one of the two basis states without sampling the full state space.
In Fig.~\ref{fig:ex2}(c), we see that all trajectories of the SSH model quickly reach one of the two momentum subspaces, even if they start in a superposition state of different momenta.
Once the trajectories have a well-defined momentum, the behavior of the system resembles that of the two-state emitter.
The convergence of quantum jump trajectories into basins associated with the steady states is a general phenomenon, which is also discussed in Ref.~\cite{SchmolkeArXiv250610873Quant-Ph2025}.

Next, we illustrate the relationship between the wall-time steady states and the jump-time steady states.
The space of wall-time steady states is spanned by
\begin{equation} \label{eq:ssh:wtss}
	\rho_k = \ketbra k k \otimes \biggl[ \frac 1 2 \mathbbm 1 + \frac{\gamma^2\, \sigma_z - 4\gamma\, (v + \ee^{\ii k} w)\, \sigma_y}{16\, (v + \ee^{\ii k} w)^2 + 2\gamma^2} \biggr] ,
\end{equation}
for $k = 0, \pi$, and that of jump-time steady states by
\begin{equation} \label{eq:ssh:jtss}
	\bar\rho_k = \ketbra k k \otimes \ketbra A A .
\end{equation}
For $v=0$, the jump-time state may exhibit an oscillatory behavior at large $n$.
Furthermore, for $w=0$, some of the coherences between the two momenta subspaces do not decay.
There are therefore additional degrees of freedom in both the wall-time steady state and the jump-time steady state, and the states listed in Eqs.~\eqref{eq:ssh:wtss} and \eqref{eq:ssh:jtss} do not span the steady state spaces any more.
We therefore focus on the case of $v$ and $w$ being nonzero and different from each other.
In that case, the state of the system approaches steady states both for the wall-time and the jump-time dynamics.

Even though individual trajectories converge into one of the momentum subspaces, the momentum distribution of the total ensemble remains constant.
The fraction of the ensemble with the momentum $k$ is $\lambda_k = \tr[ \Pi_k \rho_0 ]$, where $\rho_0$ is the initial state and $\Pi_k = \ketbra k k \otimes \mathbbm 1$ the projector onto the momentum subspace.
The states describing the system at large times or jump counts are therefore $\rho_\infty = \sum_k \lambda_k \rho_k$ and $\bar\rho_\infty = \sum_k \lambda_k \bar\rho_k$.
If the system at first is in the wall-time steady state $\rho_\infty$, its (unnormalized) state immediately after a jump is given by
\begin{equation}
	\op J \rho_\infty = \sum_k \lambda_k \op J \rho_k = \sum_k \lambda_k j_k\, \bar\rho_k ,
\end{equation}
where $j_k = \tr[ \op J \rho_k ]$.
In accordance with our discussion in Sec.~\ref{subsec:steady_states}, this new state is a jump-time steady state but it is not $\bar\rho_\infty$, because detecting the jump modifies our knowledge about the momentum distribution.

Since the jump-time steady states \eqref{eq:ssh:jtss} have no overlap with the excited states $\ket{i\mathrm B}$, the expectation value of the current between the unit cells~\cite{Asboth2016},
\begin{equation}
	C = -\ii w \bigl( \ketbra{1B}{2A} - \ketbra{2A}{1B} \bigl) ,
\end{equation}
is always zero for these states.
One may wish to find an observable in the jump-time approach that is equivalent to $C$ in the wall-time approach, where it has the expectation value
\begin{equation}
	\expval{C}_{\text{wt}} = \sum_k \frac{2 \lambda_k \gamma w\, (w + \ee^{\ii k} v)}{8\, (v + \ee^{\ii k} w)^2 + \gamma^2} .
\end{equation}
To this end, we introduce the superoperator
\begin{equation}
	\op U \rho = -\op L_0 \bigl[ (\tau_0 \Pi_0 + \tau_\pi \Pi_\pi)^{1/2}\, \rho\, (\tau_0 \Pi_0 + \tau_\pi \Pi_\pi)^{1/2} \bigr],
\end{equation}
which is designed to be bijective, preserve Hermiticity, and map $\rho_\infty$ to $\bar\rho_\infty$.
The expectation value in the jump-time steady state of the observable
\begin{equation}
	\bar C = (\op U^\dagger)^{-1} C
\end{equation}
then equals the expectation value of $C$ in the wall-time steady state.
The full expression for $\bar C$ is too complicated to include here, but it depends on all system parameters.

\section{Two-way currents} \label{sec:bidi}

\subsection{Setup} \label{subsec:bidi:setup}

In the second part of the paper, we again consider the master equation~\eqref{eq:master_equation}.
However, we now allow quantum jumps to either increase or decrease the counter, depending on their type.
For example, the counter may increase or decrease as electrons tunnel on and off a quantum dot, as illustrated in Fig.~\ref{fig:1}.
We can then write the generator as
\begin{equation} \label{eq:bidi:jump_op}
	\op L = \op L_0 + \op J_+ + \op J_- ,
\end{equation}
where $\op L_+$ and $\op L_-$ are forward and backward jump operators for particle emissions that increase the counter and particle absorptions that decrease the counter, respectively.
A jump record is now a collection 
\begin{equation} \label{eq:bidi:jump_record}
\mathcal R = (t_1, d_1; \ldots; t_N, d_N)
\end{equation}
of times $t_k$ and directions $d_k$, which indicate whether the counter increases ($d_k=+1$) or decreases ($d_k=-1$).
The total number of counted jumps $N$ is now different from the net count of emitted particles, $n_{\mathcal R} = \sum_k d_k$.

Given the initial state $\rho_0$ and the time span $[0,t]$, the probability of observing the jump record $\mathcal R$ reads~\cite{Carmichael1993, MenczelPhysRevRes2020}
\begin{equation} \label{eq:bidi:trajectory_weight}
	P_t(\mathcal R) = \tr\Bigl[ \ee^{\op L_0 (t - t_N)}\, \op J_{d_N}\, \ee^{\op L_0 (t_N - t_{N-1})}\, \cdots\, \op J_{d_1}\, \ee^{\op L_0 t_1} \rho_0 \Bigr] ,
\end{equation}
and the state conditioned on this event is
\begin{equation} \label{eq:bidi:trajectory}
	\rho_{t\, \mid\, \mathcal R} = \frac{1}{P_t(\mathcal R)}\, \ee^{\op L_0 (t - t_N)}\, \op J_{d_N}\, \ee^{\op L_0 (t_N - t_{N-1})}\, \cdots\, \mathsf J_{d_1}\, \ee^{\op L_0 t_1} \rho_0 .
\end{equation}
Averaging over all possible trajectories with the appropriate probabilities gives back the full state $\rho_t$.

As we will see, it is rather straightforward to extend our discussion of full counting statistics to two-way currents. By contrast, the theory of first-passage time distributions becomes much more involved.

\subsection{Full Counting Statistics} \label{subsec:bidi:fcs}

In full counting statistics, we again divide the full ensemble into partitions depending on the counter $n_{\mathcal R}$,
\begin{equation} \label{eq:Ttn_general}
	\rho_t(n) = \int \dD \mathcal R\, \delta_{n_{\mathcal R}, n}\, P_t(\mathcal R)\, \rho_{t\, \mid\, \mathcal R},
\end{equation}
where $n$ is any integer and $\int \dD \mathcal R$ is a shorthand notation for the sum over all trajectories,
\begin{equation}
	\int \dD \mathcal R = \sum_N \int \dd t_1\, \sum_{\mathclap{d_1 = \pm 1}} \cdots \int \dd t_N\, \sum_{\mathclap{d_N = \pm 1}} .
\end{equation}
We then introduce the $n$-resolved propagators $\op T_t(n)$ with $\rho_t(n) = \op T_t(n) \rho_0$.
They satisfy the differential equation
\begin{equation} \label{eq:Ttn_general_diffeq}
	\partial_t \op T_t(n) = \op L_0 \op T_t(n) + \op J_+ \op T_t(n-1) + \op J_- \op T_t(n+1) ,
\end{equation}
and they are generally harder to evaluate than for one-way currents.
In particular, Eq.~\eqref{eq:convolution2} does not hold here.
In Appendix~\ref{app:keldysh}, we show how Keldysh's theorem can be used to formulate a general solution to the differential equation \eqref{eq:Ttn_general_diffeq} in terms of eigenvalues and eigenvectors of the tilted generator defined below.

The full counting statistics is given by the probability distribution $P_t(n) = \tr[ \op T_t(n) \rho_0 ]$, and there is now a non-zero probability for negative values of $n$.
We can again introduce the tilted generator~\cite{BagretsPhysRevB2003, FlindtEPL2004},
\begin{equation} \label{eq:bidi:tilted}
	\op L(z) = \op L_0 + z^{-1} \op J_+ + z \op J_-,
\end{equation}
and express the generating function as
\begin{equation}
	c_t(z) = \ln \tr\bigl( \exp\bigl[\op L(z) t \bigr]\, \rho_0 \bigr).
\end{equation}
Thus, if the tilted generator has a unique eigenvalue with maximum real part, the scaled generating function becomes~\cite{LandiPRXQuantum2024}
\begin{equation} \label{eq:bidi:scaled-gf}
	c(z) = \lim_{t \to \infty} \frac 1 t\, c_t(z) = \lambda_{\mathrm{max}}^{\mathrm{re}}[ \op L(z) ].
\end{equation}
The last two equations are identical to Eqs.~\eqref{eq:fcs-gf2} and \eqref{eq:fcs-sgf} for one-way currents. Hence, if the scaled generating function is smooth around the point $z=1$, the scaled cumulants are encoded in the derivatives and they can be obtained as in Eq.~\eqref{eq:fcs_cumulants}.
Otherwise, the scaled cumulants depend on the initial condition.

\subsection{First-Passage Times} \label{subsec:bidi:fpt}

For two-way currents, the first-passage time distribution is more involved to evaluate, since a given target value $n$ can be reached multiple times on a trajectory. However, as shown in Fig.~\ref{fig:1}(b), the first-passage time is the \emph{first} time that the target is reached. Thus, when defining
the ensemble of states that reach the target  for the first time at time $t$, trajectories that have already reached it must be excluded. We write the ensemble as
\begin{equation}
	\bar\rho_n(t) = \int \dD\mathcal R\, \delta_{n_{\mathcal R}, n}\, \delta(t-t_N)\, \Theta(\mathcal R' \neq n)\, P_t(\mathcal R)\, \rho_{t\, \mid\, \mathcal R},
\end{equation}
where $\Theta(\mathcal R' \neq n)$ enforces the first-passage condition: It is one if $\sum_{k=1}^\ell d_k \neq n$ for all $\ell \leq N-1$, and zero otherwise. We then introduce the first-passage propagators $\op F_n(t)$, so that $\bar\rho_n(t) = \op F_n(t) \rho_0$.
Due to the first-passage condition, the relationship with the $n$-resolved propagators is complicated and Eqs.~\eqref{eq:fnt} and \eqref{eq:fptd_relation} no longer hold.

In order to connect the two approaches, it is useful to introduce the propagators
\begin{equation}
	\op S^m_n(t) \rho_0 = \int \dD\mathcal R\, \delta_{n_{\mathcal R}, n}\, \Theta(\mathcal R \neq m)\, P_t(\mathcal R)\, \rho_{t\, \mid\, \mathcal R},
\end{equation}
which can be understood as $n$-resolved propagators with the first-passage condition enforced as an absorbing boundary condition.
Here, $\Theta(\mathcal R \neq m)$ is one if $\sum_{k=1}^\ell d_k \neq m$ for all $\ell \leq N$, and zero otherwise.
The ensembles $\op S^m_n(t) \rho_0$ therefore include the trajectories where the counter at time $t$ reads $n$ and has not yet reached the value $m$.
These propagators vanish for $n \geq m$ if $m$ is positive, and for $n \leq m$ if $m$ is negative.
In analogy with Eq.~\eqref{eq:fnt}, we can now write
\begin{align}
	\op F_n(t) &= \op J^{\vphantom{m}}_+ \op S^n_{n-1}(t) \quad \text{if $n > 0$}, \label{eq:Fnt_general1}
\shortintertext{and}
	\op F_n(t) &= \op J^{\vphantom{m}}_- \op S^n_{n+1}(t) \quad \text{if $n < 0$}. \label{eq:Fnt_general2}
\end{align}

Similarly to the $n$-resolved propagators, the propagators with the first-passage condition satisfy the differential equation
\begin{equation}
	\partial_t \op S^m_n(t) = \op L^{\vphantom{m}}_0 \op S^m_n(t) + \op J^{\vphantom{m}}_+ \op S^m_{n-1}(t) + \op J^{\vphantom{m}}_- \op S^m_{n+1}(t)  \label{eq:Snk}
\end{equation}
if $m > 0$ and $n < m$ or if $m < 0$ and $n > m$, and $\partial_t \op S^m_n(t) = 0$ otherwise.
We then obtain the relation
\begin{equation} \label{eq:bwd_central_relation}
	\op T_t(n) = \op S^m_n(t) + \int_0^t \dd\tau\, \op T_{t-\tau}(n-m) \op F_m(\tau)
\end{equation}
by showing that the left and right-hand sides  satisfy the same initial value problem.
This relation holds for any value of $m \neq 0$.
It states that a trajectory can reach the count of $n$ either not having reached the count of $m$ by the time $t$, or by reaching the count $m$ first at the time $\tau$ and then further increasing the count by $n-m$ during the remaining time.

Now, setting $m=n$ in Eq.~\eqref{eq:bwd_central_relation} and taking the Laplace transform with respect to time, we can express the first-passage propagators in terms of the $n$-resolved propagators as
\begin{equation} \label{eq:keiji}
	\op F_n(s) = \op T_s(0)^{-1} \op T_s(n).
\end{equation}
This relationship was also found in Ref.~\cite{SaitoEPL2016}.
We can further set $n = m+1$ in Eq.~\eqref{eq:bwd_central_relation} for positive $m$, or $n = m-1$ if $m$ is negative, and after some algebra, we find
\begin{equation}
\op F_{n+1}(s) = \op F_1(s) \op F_n(s)
\end{equation}
for positive $n$, while for negative values, we have
\begin{equation}
\op F_{n-1}(s) = \op F_{-1}(s) \op F_n(s).
\end{equation}
Thus, Eq.~\eqref{eq:convolution} becomes
\begin{equation} \label{eq:bidi:convolution}
	\op F_n(s) = \op F_{\pm 1}(s)^{\abs n}
\end{equation}
for two-way currents, where the sign in $\op F_{\pm 1}$ is determined by the sign of $n$.

We see that all first-passage propagators $\op F_n(s)$ can be obtained from the one-step propagators $\op F_1(s)$ and $\op F_{-1}(s)$.
They effect the forward and backward propagation of the jump-time averaged states
\begin{equation} \label{eq:bidi:jump-time-state}
	\bar\rho_n = \op F_n(s=0)\rho_0 ,
\end{equation}
and hence play roles analogous to the tilted generator $\op L(z)$.
As before, the first-passage time distribution is the probability density $\bar P_n(t) = \tr[ \op F_n(t) \rho_0 ]$, and its Laplace transform $\bar c_n(s)$ is the corresponding moment-generating function.
Based on Eq.~\eqref{eq:bidi:convolution}, it can be written as
\begin{equation} \label{eq:bidi:fpt_gf}
	\bar c_n(s) = \tr\bigl[ \op F_{\pm 1}(s)^{\abs n} \rho_0 \bigr] ,
\end{equation}
where the sign in $\op F_{\pm 1}$ again is given by the sign of $n$.

Next, we introduce scaled generating functions~$\bar c_\pm(s)$ for the forwards and backwards distributions. If $\op F_{\pm 1}(s)$ has a unique eigenvalue with maximum absolute value, the corresponding scaled generating function becomes
\begin{equation} \label{eq:bidi:fpt_sgf}
	\bar c_{\pm}(s) = \lim_{n \to \infty} \sqrt[n]{\bar c_{\pm n}(s)} = \lambda_{\mathrm{max}}^{\mathrm{abs}}[ \op F_{\pm 1}(s) ] .
\end{equation}
This result is analogous to Eq.~\eqref{eq:fps-sgf}.
If the $\bar c_{\pm}(s)$ are smooth around $s=0$, the scaled cumulants for forwards and backwards first-passage times can be found as in Eq.~\eqref{eq:fpt_cumulants}.

In applications, the counter $n$ often has a preferred direction, which results from the underlying dynamics.
It is often convenient to choose the convention that the preferred direction is towards larger positive counts.
In the setup depicted in Fig.~\ref{fig:1}(a), for example, if a positive voltage bias is applied from source to drain, the count of tunneled electrons will eventually diverge towards positive infinity on every trajectory.
The forwards distributions are then normalized, that is, $\bar c_+(0) = 1$.
However, the backwards distributions are not normalized, since the counter might never reach a given negative target.
The value $\bar c_-(0) \in [0, 1]$ governs the total probability for the counter to ever reach a negative value:
\begin{equation} \label{eq:rare_event_prob}
	\mathcal P_{-n} = \int_0^\infty \dd t\, \bar P_{-n}(t) = \bar c_{-n}(0) \sim \bar c_-(0)^n ,
\end{equation}
where the tilde denotes asymptotic equality for large $n$.
We will discuss this topic in more detail in Sec.~\ref{subsec:fluctuation_theorem}.

To compute the first-passage time generating function in Eq.~\eqref{eq:bidi:fpt_gf} in practice, one can in principle use Eq.~\eqref{eq:keiji} to obtain the operators $\op F_{\pm 1}(s)$ as
\begin{equation} \label{eq:keiji_again}
	\op F_{\pm 1}(s) = \op T_s(0)^{-1} \op T_s(\pm 1).
\end{equation}
Here, the operators $\op T_s(n)$ are the Laplace transforms of the $n$-resolved propagators, which were defined in Eq.~\eqref{eq:Ttn_general_diffeq}.
These operators can be computed using the expression
\begin{equation} \label{eq:Tsn_in_main_text}
	\op T_s(n) = \int_0^\infty \dd t\, \ee^{-st} \oint_\gamma \frac{\dd z}{2\pi\ii} z^{n-1}\, \ee^{\op L(z) t},
\end{equation}
which we derive in Appendix~\ref{app:keldysh}.
For one-way currents and $n=0$, this expression can be shown to reduce to Eq.~\eqref{eq:convolution2}.
Here, $\mathsf L(z)$ is the tilted generator defined in Eq.~\eqref{eq:bidi:tilted}, and $\gamma$ an arbitrary contour encircling the origin of the complex plane.
However, this way of computing $\op F_{\pm 1}(s)$ may be cumbersome in practice due to the involved contour integrations and Laplace transformations.
In particular, whether the integrals in Eq.~\eqref{eq:Tsn_in_main_text} can be exchanged depends on the choice of $\gamma$, as we discuss in Appendix~\ref{app:keldysh}.
In the following section, we derive a correspondence between $\op L(z)$ and $\op F_{\pm 1}(s)$, which provides another way of obtaining $\op F_{\pm 1}(s)$ that does not require computing the $n$-resolved propagators, see Eq.~\eqref{eq:bidi:F_explicit}.
With our application in Sec.~\ref{subsec:example3}, we then provide a detailed example of how to apply this technique.

\subsection{Correspondence} \label{subsec:bidi:correspondence}

We now explore the correspondence between full counting statistics and first-passage time distributions for two-way currents. To this end, we first note that the differential equation \eqref{eq:Snk} can be alternatively written as
\begin{equation}
	\partial_t \op S^m_n(t) = \op L_0 \op S^m_n(t) + (1 - \delta_{m,n}) \op J_+ \op S^m_{n-1}(t) + \op J_- \op S^m_{n+1}(t)
\end{equation}
for positive $m$. We then set $m=1$ and apply Laplace transformations with respect to both $t$ and $n$. Specifically, we apply the operations $\sum_{n=-\infty}^\infty z^{-n}$ and $\int_0^\infty \dd t\, \ee^{-st}$ on both sides of the equation. Using the initial condition $S^1_n(0) = \delta_{n,0}$, we then arrive at the relation
\begin{align}
	 &\bigl[ \op F_1(s) - z \bigr]\, \op R_+(z,s)^{-1} = z \bigl[ \op L(z) - s \bigr] \label{eq:generalized_corr1}
\shortintertext{with}
	&\op R_+(z,s) = \sum_{n=0}^\infty z^n \int_0^\infty \dd t\, \ee^{-st}\, S^1_{-n}(t). \label{eq:relation_op1}
\end{align}
A similar calculation for $m=-1$ yields
\begin{align}
	&\bigl[ \op F_{-1}(s) - z^{-1} \bigr]\, \op R_-(z,s)^{-1} = z^{-1} \bigl[ \op L(z) - s \bigr] \label{eq:generalized_corr2}
\shortintertext{with}
	&\op R_-(z,s) = \sum_{n=0}^\infty z^{-n} \int_0^\infty \dd t\, \ee^{-st}\, S^{-1}_n(t) . \label{eq:relation_op2}
\end{align}

Equation \eqref{eq:generalized_corr1} generalizes the central result in Eq.~\eqref{eq:budini}, which we used to deduce the correspondence between generators in the first part of this paper.
The only difference is that the operator $\op T_s(0)$, which relates the right eigenvalues of the generators, must be replaced by the generalized relation operator $\op R_+(z, s)$.
For one-way currents, $\op R_+(z,s)$ reduces to $\op T_s(0)$.
Equation \eqref{eq:generalized_corr2} is a similar expression for the backwards propagator.

Based on Eqs.~\eqref{eq:generalized_corr1} and \eqref{eq:generalized_corr2}, one might think that the spectral plots of the three operators $\op L(z)$, $\op F_1(s)$ and $\op F_{-1}(s)^{-1}$ would all match.
By spectral plot of $\op F_{-1}(s)^{-1}$, we here mean a plot of the inverted eigenvalues of $\op F_{-1}(s)$, which is generally not invertible.
However, the correspondences may fail if the relation operators $\op R_+(z,s)$ or $\op R_-(z,s)$ are not well-defined.
Our next goal is therefore to better understand the relation operators.

To proceed, we use the following result, which is derived in Appendix~\ref{app:keldysh}.
First, we let $z_i(s)$ denote the values of $z$ where $\op L(z)$ has the eigenvalue $s$, and $\varphi_{ik}(s)$ and $\rho_{ik}(s)$ are the corresponding left and right eigenvectors, subject to a normalization condition described in Appendix~\ref{app:keldysh}.
Here, the index $k$ enumerates the eigenstates for eigenvalues with multiplicity larger than one.
Then, the $n$-resolved propagators for $n>0$ can be written as
\begin{equation}
	\op T_s(n) = \sum_{\mathclap{\substack{i\\z_i \to 0}}} \sum_k z_i(s)^n \rho_{ik}(s) \varphi_{ik}(s), \label{eq:keldyshs1}
\end{equation}
while for $n < 0$, they read
\begin{equation}
	\op T_s(n) = \sum_{\mathclap{\substack{i\\z_i \to \infty}}} \sum_k z_i(s)^n \rho_{ik}(s) \varphi_{ik}(s). \label{eq:keldyshs2}
\end{equation}
Here, the sums run over those indices $i$ where $z_i(s) \to 0$ or $z_i(s) \to \infty$ as $\Re(s) \to \infty$, respectively, where $\Re$ denotes the real part.

After expressing $\op S^{\pm 1}_n(s)$ in terms of the $n$-resolved and first-passage propagators by means of Eq.~\eqref{eq:bwd_central_relation}, we use Eqs.~\eqref{eq:keldyshs1} and \eqref{eq:keldyshs2} to derive the following expression:
\begin{align}
	\op R_\pm(z,s) &= \op T_s(0) + \;\,\smashoperator{\sum_{\substack{i\\z_i \to 0, \infty}}}\;\, \sum_{n=1}^{\infty}\, \Bigl[ \frac{z_i(s)}{z} \Bigr]^{\mp n} \nonumber\\ 
		& \times \sum_k \rho_{ik}(s) \varphi_{ik}(s) \bigl[ 1 - z^{\mp 1} \op F_{\pm 1}(s) \bigr].
\end{align}
Importantly, the sums over $n$ should be understood as formal power series with the value $\sum_{n=1}^\infty x^n = x / (1-x)$, which diverge only for $x = 1$.
The relation operator $\op R_+(z,s)$ is thus defined everywhere except on the curves $z = z_i(s)$ corresponding to eigenvalues that go to zero as $\Re(s) \to \infty$.
The other relation operator, $\op R_-(z,s)$ is defined everywhere except on the curves $z = z_i(s)$ for eigenvalues that diverge in that limit.
As a result, the correspondence between $\op L(z)$ and $\op F_1(s)$ holds on the curves with diverging eigenvalues, and the correspondence between $\op L(z)$ and $\op F_{-1}(s)$ holds on the curves with eigenvalues that go to zero.
As in the case of one-way currents, there may be additional exceptions to these correspondences on the line $z=0$ and on some lines of constant~$s$.

We can now summarize the generalized correspondence.
In the same way that $\op L(z)$ and $\op F_1(s)$ had matching spectral plots in the first part of the paper, the spectral plot of $\op L(z)$ now matches the union of the spectral plots of $\op F_1(s)$ and of $\op F_{-1}(s)^{-1}$.
An example can be found in Fig.~\ref{fig:ex3:splot}.
For corresponding points in the spectral plots, the left eigenvectors of the two operators are equal, and the right eigenvectors are related by the operators $\op R_\pm(z,s)$.
The correspondence can only be broken along the line $z=0$ and some lines of constant $s$.

Because of Eqs.~\eqref{eq:Fnt_general1} and \eqref{eq:Fnt_general2}, the left eigenstates corresponding to the zero eigenvalues of $\op F_{\pm 1}(s)$ are the same as those of $\op J_\pm$.
In analogy to our discussion around Eq.~\eqref{eq:uni:F_explicit}, all eigenvalues and eigenstates of $\op F_{\pm 1}(s)$ can be determined from those of $\op L(z)$.
It is thus possible to construct $\op F_{\pm 1}(s)$ from the eigenvalues and eigenstates of $\op L(z)$ by using the diagonalization
\begin{equation} \label{eq:bidi:F_explicit}
	\op F_{\pm 1}(s) = \op X_\pm(s)^{-1}\, \op \Lambda_\pm(s)\, \op X_\pm(s) ,
\end{equation}
where, as in Eq.~\eqref{eq:uni:F_explicit}, $\Lambda_\pm(s)$ are diagonal matrices and $\op X_\pm(s)$ basis transformation matrices.
The diagonal matrix $\Lambda_+(s)$ [$\op \Lambda_-(s)$] contains the eigenvalues of $\op F_1(s)$ [$\op F_{-1}(s)$], which are those $z_i(s)$ that go to zero [diverge] as $\Re(s) \to \infty$, and a zero eigenvalue whose multiplicity equals the rank deficiency of $\op J_+$ [$\op J_-$].
The rows of the matrices $\op X_\pm(s)$ are the corresponding left eigenstates of $\op J_\pm$ or of $\op L(z)$.
This technique for obtaining $\op F_{\pm 1}(s)$ is often easier than using Eqs.~\eqref{eq:keiji_again} and \eqref{eq:Tsn_in_main_text}, which rely on first finding the $n$-resolved propagators, as we will illustrate with the application in Sec.~\ref{subsec:example3}.

\subsection{Steady States, Scaled Cumulants, and Fluctuation Theorems} \label{subsec:fluctuation_theorem}

As we have found, the correspondence from the first part of the paper has a straightforward generalization to two-way currents.
For one-way currents, we have already discussed its implications on steady states, observables, large-deviation functions, and scaled cumulants.
These results also largely apply to the more general situation of two-way currents, and we briefly discuss them now.

Depending on the specifics of the system, it is possible that its trajectories reach every possible count with probability one (e.g., symmetric random walk), that the counter tends towards infinity (either positive or negative) with probability one, that the counter is bounded to a finite interval, or anything in between.
To simplify the discussion, we focus in the remainder on the typical case, where the current has a preferred direction that does not depend on the initial condition.
Specifically, we assume that the counter approaches positive infinity with probability one.
In that case, the spectral plot of $\op L(z)$ around the important point $(z=1,s=0)$ fully matches the spectral plot of $\op F_1(s)$ around that point.

We can then draw the following conclusions, which are in full analogy with the case of one-way currents:
\begin{enumerate}[(1), leftmargin=*]
\item The wall-time steady states, defined by $\op L \rho = 0$, are in one-to-one correspondence with the forward jump-time steady states, given by $\op F_1 \bar\rho = \bar\rho$. The relation operator $\op R_+(1,0)$ is an isomorphism between the two steady state spaces, but it does not necessarily connect $\rho_\infty$ and $\bar\rho_\infty$ for a given initial condition.
\item The scaled generating functions $c(z)$ and $\bar c_+(s)$ are inverse functions.
	In the case of unique steady states, the cumulant relations \eqref{eq:cumulant_relations1} and \eqref{eq:cumulant_relations2} hold.
\item If the steady states are not unique, or if the observation times are shorter than the lifetimes of metastable states, the cumulant relations can be violated. This phenomenon can be used to detect metastability.
\end{enumerate}

The scaled generating function $c(z)$ is in correspondence with $\bar c_+(s)$ locally around $z=1$.
Globally, it may consist of parts corresponding to $\bar c_+(s)$ and parts corresponding to $\bar c_-(s)$.
Typically, as seen in Fig.~\ref{fig:ex3:splot} for the driven qubit that we consider in the next section, $c(z)$ has a decreasing part and an increasing part, which correspond to $\bar c_+(s)$ and $\bar c_-(s)$, respectively.
The large-deviation function $I(j)$ of the full counting statistics then corresponds to the large-deviation function of the forwards first-passage time, $\bar I_+(\tau)$, for $j > 0$, and to $\bar I_-(\tau)$ for $j < 0$.
Specifically, they satisfy the relation~\cite{GingrichPhysRevLett2017}
\begin{equation}
	I(j) = \abs{j}\, \bar I_\pm(1 / \abs{j}),
\end{equation}
where the sign used in $\bar I_\pm$ is the same as the sign of $j$.

Because of the shape of $c(z)$, it must have an additional zero for some value of $z > 1$, which we denote by $r$.
The Perron-Frobenius eigenvalue of the quantum channel $\op F_{-1}(0)$ must be $1/r$, and we therefore have $\bar c_-(0) = 1 / r$.
Following Sec.~\ref{subsec:bidi:fpt}, we then obtain the relation
\begin{equation} \label{eq:bidi:fthm}
	\mathcal P_{-n} / \mathcal P_n \sim r^{-n} ,
\end{equation}
where $\mathcal P_n = \int_0^\infty \dd t\, \bar P_n(t)$, which is one for $n > 0$, and the tilde denotes asymptotic equivalence for large $n$.

The relation \eqref{eq:bidi:fthm} is a fluctuation theorem for the integrated first-passage time distributions $\mathcal P_n$, which are the total probabilities for the counter to ever reach the respective values $n$.
The fluctuation theorem states that it is exponentially unlikely for the counter to reach large negative values.
The parameter $r$, which is formally defined as the smallest value of the counting field larger than one such that the tilted generator has a zero eigenvalue, controls the strength with which these probabilities are exponentially suppressed.
It therefore quantifies the strength of the bias towards positive counts: small $r \gtrsim 1$ correspond to small bias and large $r$ to large bias.
The fluctuation theorem provides a direct connection between the probability of rare events, where the counter reaches large negative values, and the spectrum of $\op L(z)$.
We note that it holds only asymptotically in the limit of large~$n$.
For small values of $n$, the probabilities $\mathcal P_{-n}$ may depend on the initial conditions, as we will see in the following section.

To obtain another perspective on the meaning of the parameter $r$, let us assume that the underlying dynamics is microreversible or, equivalently, that the generator $\op L$ obeys a local detailed balance condition.
In this case, the scaled generating function $c(z)$ obeys the Gallavotti-Cohen symmetry~\cite{LebowitzJStatPhys1999, AndrieuxJStatMech2007, ManzanoPhysRevB2014}
\begin{equation}
	c(z) = c(\ee^{\mathcal A} / z) ,
\end{equation}
where $\mathcal A$ is the thermodynamic affinity associated with the current.
In this situation, we have $r = \exp(\mathcal A)$; the strength of the bias towards positive counts is therefore directly related to the affinity.
The fluctuation theorem can then be rewritten in the familiar form
\begin{equation} \label{eq:bidi:fthm2}
	\mathcal P_{-n} / \mathcal P_n \sim \ee^{-\mathcal A n}.
\end{equation}
For dynamics that does not satisfy local detailed balance, the Gallavotti-Cohen symmetry is generally not satisfied.
In this case, the parameter $\ln(r)$ can be viewed as a generalized thermodynamic affinity.
Note however that such systems might not admit a thermodynamically consistent description.

\subsection{Application: Driven Qubit} \label{subsec:example3}

As our last application, we consider the qubit in Fig.~\ref{fig:examples}(c), which is in contact with a thermal environment and subject to an oscillatory driving field.
At zero detuning of the levels, the Hamiltonian of the qubit in a rotating frame reads
\begin{equation}
    H = \hbar\Omega\, \sigma_x.
\end{equation}
The environment causes spontaneous emissions and absorptions, which are described by the Lindblad operators $L_\up = \sigma_+$ and $L_\dwn = \sigma_-$ and the rates $\gamma_\dwn > \gamma_\up > 0$.
We assume that the driving strength $\Omega$ is non-zero, because otherwise the two Lindblad operators would induce jumps in a strictly alternating way, and the counter $n$ could never reach values larger than one or smaller than minus one.

To keep track of the energy that is dissipated from the qubit into the environment, we define
\begin{align}
	\op J_+ \rho &= \gamma^{\vphantom\dagger}_\dwn\, L^{\vphantom\dagger}_\dwn \rho L_\dwn^\dagger
\shortintertext{and}
	\op J_- \rho &= \gamma^{\vphantom\dagger}_\up\, L^{\vphantom\dagger}_\up \rho L_\up^\dagger.
\end{align}
According to the second law of thermodynamics, the dissipated energy can only increase on average~\cite{SpohnJMathPhys1978}.
Thus, while the net dissipated energy can temporarily become negative on some trajectories~\cite{SaitoEPL2016}, it must eventually increase towards infinity.
This effect can be seen in Fig.~\ref{fig:1}(b), displaying randomly generated trajectories for this setup.

\begin{figure}[t]
	\centering
	\includegraphics[scale=1]{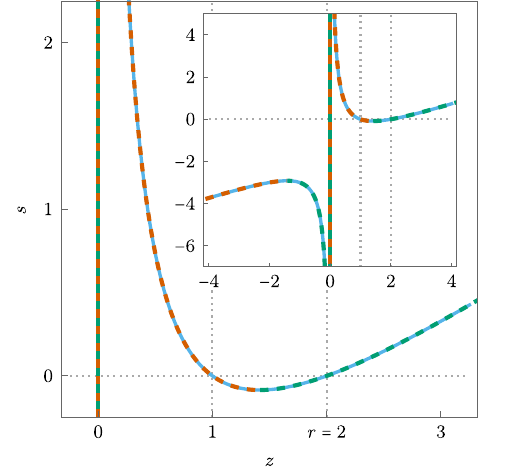}
	\caption{
		Driven qubit.
        We show spectral plots of the operators $\op L(z)$ (solid blue), $\op F_1(s)$ (dashed orange), and $\op F_{-1}(s)^{-1}$ (dashed green).
        Note that at $z=0$, there is no vertical blue line.
		The inset shows a zoomed-out version of the plot.
		Dotted lines highlight $s=0$ and the values $z = 1$ and $z = r = \gamma_\dwn / \gamma_\up$, where $\op L(z)$ has zero eigenvalues.
		This figure illustrates that, for two-way currents, the spectral plot of the full counting statistics generator matches the combined spectral plots of the forward and backward first-passage propagators.
		Here, we have used arbitrary units with $\Omega = 10$, $\gamma_\dwn = 2$ and $\gamma_\up = 1$.
	}
	\label{fig:ex3:splot}
\end{figure}

\begin{figure}[t]
	\centering
	\includegraphics[scale=1]{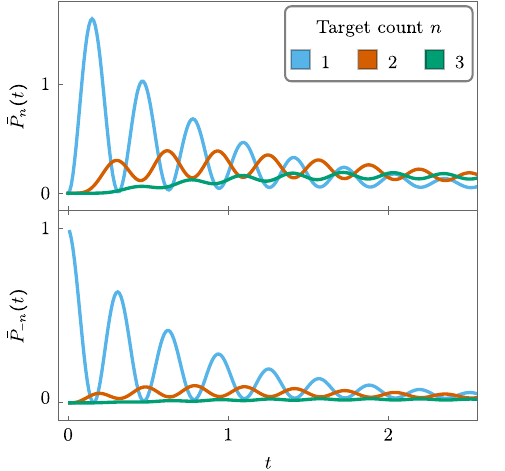}
	\caption{
		First-passage time distributions for the driven qubit. We show the distributions $\bar P_n(t)$ and $\bar P_{-n}(t)$ for small values of $n$, using the same parameters as in Fig.~\ref{fig:ex3:splot}.
		The results show an oscillatory behavior, which can be traced back to the Rabi oscillations of the driven qubit modulating the probabilities of absorption and emission over time.
	}
	\label{fig:ex3:fptd}
\end{figure}

We now show how our correspondence \eqref{eq:bidi:F_explicit} can be used for finding the first-passage propagators $\op F_{\pm 1}(s)$. In practice, it is useful to express the superoperators $\op L_0$ and $\op J_{\pm}$ in the basis $(\sigma_y / 2, \sigma_z / 2, \mathbbm 1 / 2)$, where they read
\begin{equation}
	\op J_+ = \begin{pmatrix}
			0 & 0 & 0 \\
			0 & -\frac{\gamma_\dwn}{2} & -\frac{\gamma_\dwn}{2} \\
			0 & \frac{\gamma_\dwn}{2} & \frac{\gamma_\dwn}{2} \\
		\end{pmatrix} 
        , \quad
	\op J_- = \begin{pmatrix}
			0 & 0 & 0 \\
			0 & -\frac{\gamma_\up}{2} & \frac{\gamma_\up}{2} \\
			0 & -\frac{\gamma_\up}{2} & \frac{\gamma_\up}{2} \\
		\end{pmatrix},
\end{equation}
and
\begin{equation}
	\op L_0 = \begin{pmatrix}
			-\frac{\gamma_\dwn + \gamma_\up}{2} & -2\Omega & 0 \\
			2\Omega & -\frac{\gamma_\dwn + \gamma_\up}{2} & -\frac{\gamma_\dwn - \gamma_\up}{2} \\
			0 & -\frac{\gamma_\dwn - \gamma_\up}{2} & -\frac{\gamma_\dwn + \gamma_\up}{2}
		\end{pmatrix}.
\end{equation}
Note that we ignore the $x$-component of the Bloch vector, since it fully decouples from the rest of the dynamics.

Since both $\op J_+$ and $\op J_-$ have double zero eigenvalues, also $\op F_1(s)$ and $\op F_{-1}(s)$ have double zero eigenvalues.
The corresponding left eigenvectors, which are shared between $\op J_+$ and $\op F_1(s)$ and between $\op J_-$ and $\op F_{-1}(s)$, are
\begin{align}
	&\varphi_{0,1} = (1, 0, 0), \quad \varphi_{0,2} = (0, 1, 1),
\shortintertext{and}
	&\bar\varphi_{0,1} = (1, 0, 0), \quad \bar\varphi_{0,2} = (0, -1, 1).
\end{align}
To determine the third and final eigenvalues of $\op F_1(s)$ and $\op F_{-1}(s)$, we solve the equation $\det[\op L(z) - s] = 0$ for $z$ and obtain the solutions
\begin{align}
	z_1(s) &= \frac{\zeta(s) - d(s)}{2\gamma_\up}
\shortintertext{and}
	\bar z_1(s) &= \frac{\zeta(s) + d(s)}{2\gamma_\up},
\end{align}
where we have defined
\begin{equation}
	\zeta(s) = (2s + \gamma_\dwn + \gamma_\up) \biggl[ 1 + \frac{s(s + \gamma_\dwn + \gamma_\up)}{4\Omega^2} \biggr]
\end{equation}
and 
\begin{equation}
	d(s) = [\zeta(s)^2 - 4\gamma_\dwn \gamma_\up]^{1/2} .
\end{equation}
The corresponding left eigenvectors of $\op L(z_1(s))$ and $\op L(\bar z_1(s))$ read
\begin{align}
	\varphi_1(s) &= \biggl( \frac{s(s + \gamma_\dwn + \gamma_\up) / \Omega}{d(s) + \gamma_\dwn - \gamma_\up}, f(s) [d(s) + \gamma_\up - \gamma_\dwn], 1 \biggr)
\intertext{and}
	\bar\varphi_1(s) &= \biggl( \frac{s(s + \gamma_\dwn + \gamma_\up) / \Omega}{d(s) + \gamma_\up - \gamma_\dwn}, f(s) [d(s) + \gamma_\dwn - \gamma_\up], -1 \biggr),
\end{align}
where we have introduced the function
\begin{equation}
	f(s) = \frac{4\Omega^2 (2s + \gamma_\dwn + \gamma_\up)}{[ s(s + \gamma_\dwn + \gamma_\up) + 8\Omega^2 ] (2s + \gamma_\dwn + \gamma_\up)^2 + 64\Omega^4} .
\end{equation}

As $s$ approaches infinity, the eigenvalues $z_1(s)$ and $\bar z_1(s)$ go to zero and infinity, respectively.
Therefore, $z_1(s)$ is the missing eigenvalue of $\op F_1(s)$ and $\bar z_1(s)^{-1}$ the one of $\bar F_{-1}(s)$.
We can then write down the first-passage propagators using Eq.~\eqref{eq:bidi:F_explicit}.
To this end, we define $\op X(s)$ as a matrix consisting of the rows $\varphi_{0,1}$, $\varphi_{0,2}$, and $\varphi_1(s)$. We then find
\begin{align}
	\op F_1(s) &= \op X(s)^{-1} \begin{pmatrix} 0 & 0 & 0 \\ 0 & 0 & 0 \\ 0 & 0 & z_1(s) \end{pmatrix} \op X(s) \nonumber\\
		&= \frac{z_1(s)}{f_2(s) [d(s) + \gamma_\up - \gamma_\dwn]} \begin{pmatrix}
				0 & 0 & 0 \\
				\text{--} & \varphi_1(s) & \text{--} \\
				\text{--} & -\varphi_1(s) & \text{--}
			\end{pmatrix},
\end{align}
and
\begin{equation}
	\op F_{-1}(s) = \frac{\bar z_1(s)^{-1}}{f_2(s) [d(s) + \gamma_\dwn - \gamma_\up]} \begin{pmatrix}
				0 & 0 & 0 \\
				\text{--} & \bar\varphi_1(s) & \text{--} \\
				\text{--} & \bar\varphi_1(s) & \text{--}
			\end{pmatrix}
\end{equation}
can be obtained in a similar way.
Here, the notation implies that the matrices consist of the indicated row vectors.

Figure~\ref{fig:ex3:splot} shows spectral plots of the operators $\op L(z)$, $\op F_1(s)$ and $\op F_{-1}(s)^{-1}$ for the qubit.
It illustrates how, except for the zero eigenvalues of $\op F_{\pm 1}(s)^{\pm 1}$, the spectral plot of $\op L(z)$ comprises those of the two other operators.

The dominant eigenvalue of $\op L(z)$ has zeros at $z=1$ and at $z = \gamma_\dwn / \gamma_\up = r$.
Following our general theory, the probability to ever reach a count of $-n$ therefore scales as
\begin{equation}
	\mathcal P_{-n} \sim r^{-n} = \bigl( \gamma_\up / \gamma_\dwn \bigr)^{-n}
\end{equation}
for large $n$.
For a given initial state $\rho_0$, the exact value is given by $\mathcal P_{-n} = \tr[ \op F_{-1}(0)^n \rho_0 ]$.
Taking the initial state to be the ground state $\ket 0$, we obtain the exact result
\begin{equation}
	\mathcal P_{-n} = \frac{\gamma_\dwn (\gamma_\dwn + \gamma_\up) + 4\Omega^2}{\gamma_\up (\gamma_\dwn + \gamma_\up) + 4\Omega^2} \Bigl( \frac{\gamma_\up}{\gamma_\dwn} \Bigr)^{-n} .
\end{equation}

Because the system is in the ground state, there is a large probability to have a spontaneous absorption initially.
That effect enhances $\mathcal P_{-n}$ for small $n$, but the influence of the initial condition becomes increasingly unimportant for larger $n$.
In Fig.~\ref{fig:ex3:fptd}, we show the first-passage time distributions for this example with $-3 \leq n \leq 3$, which we computed by first using Eq.~\eqref{eq:bidi:fpt_gf} to obtain the moment-generating functions analytically, and then applying numerical inverse Laplace transforms.
This procedure is more efficient than sampling the distribution from numerical simulations of the dynamics.
The figure shows that $\bar P_{-1}(t)$ is finite at $t=0$ because of the initial condition, but $\bar P_1(t)$ and all other distributions are initially zero.
The oscillatory behavior arises from Rabi oscillations, where an emission or absorption event is most likely at the times where the system is close to its excited or ground state, respectively.

\section{Conclusion and perspectives} \label{sec:perspectives}

We have developed a general theoretical framework that connects full counting statistics and first-passage time distributions for quantum systems described by Markovian Lindblad master equations.
By deriving ensemble relations between the two descriptions, we have provided a unified perspective on time-integrated and time-resolved fluctuations in open quantum systems, and we have established a connection between the long-time and the large-$n$ steady states.
We have analyzed the role of metastability, where long-lived intermediate states give rise to nontrivial correlations.
Furthermore, we have extended the scope of fluctuation theorems, which are typically formulated for time-integrated quantities, to first-passage time distributions.

Our framework can be extended in several ways.
One natural direction would be to explore more general counting processes, where the counter may take steps of arbitrary size instead of taking only integer steps.
While the asymptotic correspondence continues to hold in this case~\cite{GingrichPhysRevLett2017}, extending the finite-time framework is more difficult due to the possibility of ``overshooting'' the first-passage target value.
Still, we expect that our results would also hold in this generalized scenario.
We further expect that our framework could be generalized to diffusive processes like those considered in Ref.~\cite{KewmingPhysRevA2024}, and that our physical results would still hold there as well.

An interesting area of research concerns first-passage times for non-Markovian systems~\cite{GuerinNature2016, AkimotoPhysRevE2020}.
While we have considered memoryless evolution, many realistic systems, including classical systems with anomalous diffusion~\cite{ScherPhysRevB1975, HoflingRepProgPhys2013} and quantum systems with strong reservoir couplings~\cite{Breuer2002, LeggettRevModPhys1987, ZhangPhysRevLett2012, BreuerRevModPhys2016, deVegaRevModPhys2017, TalknerRevModPhys2020}, are non-Markovian.
Extending the ensemble relations and fluctuation theorems to non-Markovian systems would improve our understanding of memory effects in quantum stochastic processes and potentially reveal new fluctuation relations~\cite{FlindtPhysRevLett2008, FlindtPhysRevB2010}.

Another promising avenue concerns exceptional points, where both eigenvalues and eigenvectors of a non-Hermitian evolution operator coalesce~\cite{HeissJPhysAMathTheor2012, MiriScience2019, OzdemirNatMater2019, MingantiPhysRevA2019, KhandelwalPRXQuantum2021, LinNatCommun2025}.
For the full counting statistics, exceptional points can manifest as non-analyticities in the cumulant generating function, signaling abrupt changes in the fluctuations \cite{GarrahanPhysRevLett2010, PavlovPhysRevB2025}.
Likewise, first-passage time distributions, being sensitive to the spectral structure of the underlying generator, could exhibit distinctive features near exceptional points.
Investigating how exceptional points influence the ensemble relations may reveal aspects of  geometry and topology.

An important class of systems, where our framework may find direct application, is that of Gaussian states, which occur in quantum optics and bosonic quantum systems~\cite{Gardiner2004, WeedbrookRevModPhys2012}.
For Gaussian states, the dynamics remains analytically tractable, allowing for explicit calculations of the full counting statistics~\cite{KansanenSciPostPhys2025} and possibly also the first-passage time distribution.
Incorporating the ensemble relations into this context could yield insights about the time-resolved structure of fluctuations in continuous-variable systems.
Moreover, Gaussian states are routinely prepared and measured in the laboratory, making them ideal platforms for experimental verification of the connections between time-integrated and first-passage observables established in this work.

Experimentally, the results presented here offer concrete predictions for a variety of platforms, including superconducting circuits, mesoscopic conductors, and quantum optical systems.
Based on the progress in the real-time tracking of quantum trajectories~\cite{GleyzesNature2007, KubanekNature2009, SayrinNature2011, VijayPhysRevLett2011, MurchNature2013, DelteilPhysRevLett2014, ChantasriPhysRevX2016, WeberComptesRendusPhys2016, KurzmannPhysRevLett2019}, experimental tests of the relations and fluctuation theorems derived in this work may soon become feasible.

Finally, the connections between full counting statistics, first-passage times, and metastability suggest intriguing links to dynamical phase transitions~\cite{HeylRepProgPhys2018, MacieszczakPhysRevRes2021}.
Exploring how ensemble relations behave across different dynamical regimes could yield a more complete picture of nonequilibrium quantum phenomena and further illuminate the role of fluctuations at the nanoscale.
Our work provides a versatile framework for further investigations of quantum fluctuations, with implications across quantum thermodynamics, quantum transport, and quantum computing.

\begin{acknowledgments}
We thank C.~Jarzynski, F.~Schmolke, J.~Chiel and K.~Saito for useful discussions.
The trajectories in Figs.~\ref{fig:1}, \ref{fig:ex1} and \ref{fig:ex2} were generated using QuTiP~\cite{JohanssonComputPhysCommun2012, JohanssonComputPhysCommun2013, LambertPhysicsReports2026}.
P.~M.\ performed this work as an International Research Fellow of the Japan Society for the Promotion of Science (JSPS).
We acknowledge the support from the Research Council of Finland through the Finnish Centre of Excellence in Quantum Technology (Grant No.\ 352929), and the Japan Society for the Promotion of Science through an Invitational Fellowship for Research in Japan.
F.~N.\ is supported in part by the Japan Science and Technology Agency (JST) [via the CREST Quantum Frontiers program Grant No.\ JPMJCR24I2, the Quantum Leap Flagship Program (Q-LEAP), and the Moonshot R\&D Grant Number JPMJMS2061], and the Office of Naval Research (ONR) Global (via Grant No.\ N62909-23-1-2074).
C.~G.\ was partially supported by RIKEN Incentive Research Projects.
\end{acknowledgments}

\appendix
\makeatletter
\renewcommand{\p@subsection}{\Alph{section}}
\makeatother

\begin{table*}[p]
	\centering
	\renewcommand{\arraystretch}{1.35}
	\setlength\tabcolsep{0pt}
	\begin{tabular*}{\linewidth}{@{\extracolsep{\fill}}llll@{}}
		\toprule
		\bf Symbol & \bf Description & \multicolumn{2}{l}{\bf Defined in} \\
		\cmidrule{3-4}
		&& \emph{One-way} & \emph{Two-way} \\
		\midrule
		$c(z)$ & Scaled generating function of the full counting statistics & Eq.~\eqref{eq:fcs-sgf} & Eq.~\eqref{eq:bidi:scaled-gf} \\
		$c_t(z)$ & Generating function of the full counting statistics & Eq.~\eqref{eq:fcs-gf} & Eq.~\eqref{eq:fcs-gf} \\
		$\bar c(s)$, $\bar c_\pm(s)$ & Scaled generating function (\emph{two-way}: functions) of the first-passage times & Eq.~\eqref{eq:fps-sgf} &  Eq.~\eqref{eq:bidi:fpt_sgf} \\
		$\bar c_n(s)$ & Generating function of the first-passage times & Eq.~\eqref{eq:fps-gf} & Eq.~\eqref{eq:fps-gf} \\
		$\op F_1$ & Evolution operator for jump-time averaged states & Eq.~\eqref{eq:jump-time-state-evo} & Eq.~\eqref{eq:bidi:jump-time-state} \\
		$\op F_n(t)$, $\op F_n(s)$ & First-passage propagator, its Laplace transform & Eq.~\eqref{eq:fnt} & Eqs.~\eqref{eq:Fnt_general1}, \eqref{eq:Fnt_general2} \\
		$\gamma_{ij}$ & Transition rate in classical rate equation & Eq.~\eqref{eq:classical_ME} & Eq.~\eqref{eq:classical_ME} \\
		$\gamma_\mu$ & Relaxation rate in Lindblad equation & Eq.~\eqref{eq:lindblad} & Eq.~\eqref{eq:lindblad} \\
		$H$ & System Hamiltonian & Eq.~\eqref{eq:lindblad} & Eq.~\eqref{eq:lindblad} \\
		$I(j)$ & Large-deviation function of the full counting statistics & Eq.~\eqref{eq:ldf_definition_j} & Sec.~\ref{subsec:fluctuation_theorem} \\
		$\bar I(\tau)$, $\bar I_\pm(\tau)$ & Large-deviation function (\emph{two-way}: functions) of the first-passage times & Eq.~\eqref{eq:ldf_definition_tau} & Sec.~\ref{subsec:fluctuation_theorem} \\
		$j$ & Average output current, $j = n / t$ & Sec.~\ref{subsec:cumulant_relations} & Sec.~\ref{subsec:fluctuation_theorem} \\
		$\op J$, $\op J_\pm$ & (Forward and backward) jump operators & Eq.~\eqref{eq:jump_op} & Eq.~\eqref{eq:bidi:jump_op} \\
		$\op L$ & Generator of the time evolution (Lindbladian or rate matrix) & Eq.~\eqref{eq:master_equation} & Eq.~\eqref{eq:master_equation} \\
		$\op L_0$ & Generator of the time evolution without the counted transitions & Eq.~\eqref{eq:jump_op} & Eq.~\eqref{eq:bidi:jump_op} \\
		$L_\mu$ & Lindblad jump operator & Eq.~\eqref{eq:lindblad} & Eq.~\eqref{eq:lindblad} \\
		$\op L(z)$ & Tilted generator, generator of the time evolution with counting field & Eq.~\eqref{eq:tilted_gen} & Eq.~\eqref{eq:bidi:tilted} \\
		$n$ & Net number of emissions (\emph{one-way}: equal to total jump count) & Sec.~\ref{subsec:uni:setup} & Sec.~\ref{subsec:bidi:setup} \\
		$N$ & Total jump count & --- & Sec.~\ref{subsec:bidi:setup} \\
		$\varphi_i$, $\bar\varphi_i$ & Basis of left wall-time eigenstates, basis of left jump-time eigenstates & Sec.~\ref{subsec:steady_states} & Sec.~\ref{subsec:fluctuation_theorem} \\
		$\mathcal P_n$ & Total probability for the counter to ever reach the value $n$ & --- & Eq.~\eqref{eq:rare_event_prob} \\
		$P_t(n)$ & Full counting statistics probability distribution & Eq.~\eqref{eq:fcs} & Sec.~\ref{subsec:bidi:fcs} \\
		$P_t(\mathcal R)$ & Probability of obtaining jump record [\emph{one-way}: written as $P_t(t_1 \cdots t_n)$] & Eq.~\eqref{eq:trajectory_weight} & Eq.~\eqref{eq:bidi:trajectory_weight} \\
		$\bar P_n(t)$ & First-passage time distribution & Eq.~\eqref{eq:fptd} & Sec.~\ref{subsec:bidi:fpt} \\
		$\rho$, $\rho_i$, $\rho_\infty$ & Wall-time steady states (generic, basis, dynamically obtained) & Sec.~\ref{subsec:steady_states} & Sec.~\ref{subsec:fluctuation_theorem} \\
		$\rho_t$ & System state (density matrix or probability vector, wall-time averaged) & Sec.~\ref{subsec:uni:setup} & Sec.~\ref{subsec:bidi:setup} \\
		$\rho_{t\, \mid\, \mathcal R}$ & System state conditioned on jump record (\emph{one-way}: written as $\rho_{t\, \mid\,  t_1 \,\cdots\, t_n}$) & Eq.~\eqref{eq:trajectory} & Eq.~\eqref{eq:bidi:trajectory} \\
		$\bar\rho$, $\bar\rho_i$, $\bar\rho_\infty$ & Jump-time steady states (generic, basis, dynamically obtained) & Sec.~\ref{subsec:steady_states} & Sec.~\ref{subsec:fluctuation_theorem} \\
		$\bar\rho_n$ & Jump-time averaged state & Eq.~\eqref{eq:jump-time-state} & Eq.~\eqref{eq:bidi:jump-time-state} \\
		$\mathcal R$ & Jump record [\emph{one-way}: written as $(t_1 \cdots t_n)$] & --- & Eq.~\eqref{eq:bidi:jump_record} \\
		$\op R_\pm(z, s)$ & Relation operators in two-way correspondence & --- & Eqs.~\eqref{eq:relation_op1}, \eqref{eq:relation_op2} \\
		$s$ & Laplace parameter (counting field for $t$), or eigenvalue of $\op L(z)$ & Sec.~\ref{subsec:uni:fpt} & Sec.~\ref{subsec:bidi:fpt} \\
		$\op S^m_n(t)$ & $n$-resolved propagators with boundary conditions & --- & Eq.~\eqref{eq:Snk} \\
		$t$ & Time (as shown by a clock on the wall, ``wall time'') & Sec.~\ref{subsec:uni:setup} & Sec.~\ref{subsec:bidi:setup} \\
		$\tau$ & Average first-passage time, $\tau = t / n$ & Sec.~\ref{subsec:cumulant_relations} & Sec.~\ref{subsec:fluctuation_theorem} \\
		$\op T_t(n)$, $\op T_s(n)$ & $n$-resolved propagators, their Laplace transforms & Eq.~\eqref{eq:n-resolved-prop} & Eq.~\eqref{eq:Ttn_general_diffeq} \\
		$z$ & Counting field for $n$, or eigenvalue of $\op F_n(s)$ & Sec.~\ref{subsec:uni:fcs} & Sec.~\ref{subsec:bidi:fcs} \\
		$\expval{n^k}_t$, $\cumulant{n^k}_t$ & Moments and cumulants of the full counting statistics & Eq.~\eqref{eq:fcs-cumulants} & Eq.~\eqref{eq:fcs-cumulants} \\
		$\cumulant{j^k}$ & Scaled cumulants of the full counting statistics & Eq.~\eqref{eq:fcs_cumulants} & Eq.~\eqref{eq:fcs_cumulants} \\
		$\expval{t^k}_n$, $\cumulant{t^k}_n$ & Moments and cumulants of the first-passage times & Eq.~\eqref{eq:fpt-cumulants} & Eq.~\eqref{eq:fpt-cumulants} \\
		$\cumulant{\tau^k}$ & Scaled cumulants of the first-passage times (\emph{two-way}: forwards) & Eq.~\eqref{eq:fpt_cumulants} & Eq.~\eqref{eq:fpt_cumulants} \\
		\bottomrule
	\end{tabular*}
	\caption{
		Alphabetical list of important symbols, their meanings, and where they are defined for one-way currents (Sec.~\ref{sec:counting}) and two-way currents (Sec.~\ref{sec:bidi}).
		The definitions for two-way currents generalize the definitions for one-way currents.
	}
	\label{tab:notation}
\end{table*}

\begin{figure*}[!t]
	\centering
    \includegraphics[scale=1]{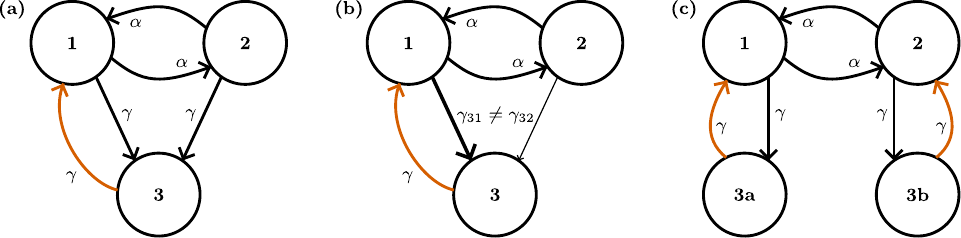}
	\caption[]{
		Examples discussed in Appendix~\ref{app:quick_example}.
		\begin{enumerate*}[(a)]
		\item
			Three-state system following a classical master equation.
			The jump rates between states ``1'' and ``2'' are $\alpha$, and the jump rates from these states to state ``3'' are $\gamma$.
			Only jumps from ``3'' back to ``1'', also with rate $\gamma$, are visible.
			Since the transitions between ``1'' and ``2'' do not affect the times where jumps from ``3'' to ``1'' occur, and since they also do not influence the jump-time averaged states, the first-passage propagator does not depend on $\alpha$.
		\item
			Modification where the jump rate from ``1'' to ``3'' ($\gamma_{31}$) and the jump rate from ``2'' to ``3'' ($\gamma_{32}$) differ, in which case the first-passage time distribution depends on $\alpha$.
		\item
			Modification where the information on whether we came from state ``1'' or ``2'' is retained, and the visible jumps go back to the original state.
			The modification results in jump-time averaged states that depend on $\alpha$.
		\end{enumerate*}
		In both examples (b) and (c), the first-passage propagator depends on $\alpha$, and the tilted generator is fully determined by the first-passage propagator.
	}
	\label{fig:appc}
\end{figure*}

\section{Spectral Properties and Dark States} \label{app:dark_states}

Here, we demonstrate the following two statements:
\begin{enumerate}[(1), leftmargin=*]
\item A state $\rho$ is a dark state if and only if $\op L_0 \rho = 0$.
\item If $\mathsf L_0$ has one or more zero eigenvalues, the corresponding eigenspace is spanned by dark states.
\end{enumerate}
As in the main text, we have written the generator as $\op L = \op L_0 + \op J$, where $\op J$ is the jump operator and $\op L_0$ denotes the rest. In general, the jump operator can be written as
\begin{equation}
	\op J \rho = \sum_\mu \gamma_\mu \eta_\mu\, L_\mu \rho L_\mu^\dagger ,
\end{equation}
where $\gamma_\mu$ and $L_\mu$ are the rates and Lindblad operators from Eq.~\eqref{eq:lindblad}, and $\eta_\mu \in [0, 1]$.
We can write $\op L_0$ as
\begin{equation} 
	\op L_0 \rho = \op L' \rho - \acomm{D}{\rho} ,
\end{equation}
where
\begin{equation}
D = \frac 1 2 \sum_\mu \gamma_\mu \eta_\mu\, L_\mu^\dagger L_\mu .
\end{equation}

The free generator $\op L'$ has the Lindblad form and is therefore completely positive and trace preserving.
We define a dark state to be a physical state $\rho$ with $\op L' \rho = 0$ and $\op J \rho = 0$, where ``physical state'' refers to a positive semi-definite operator with unit trace.
The three superoperators $\op L$, $\op L_0$ and $\op L'$ are all generators of completely positive time evolution. The spectra of all three are thus contained in the closed left half-plane.
Since $\op L$ and $\op L'$ are trace-preserving, they each have at least one zero eigenvalue.

To proceed, we first show the auxiliary statement:
\begin{enumerate}[(1), leftmargin=*, start=3]
\item If $\op L_0$ has an eigenvalue with zero real part, the corresponding eigenoperator is also an eigenoperator of $\op L$ and $\op L'$ with the same eigenvalue, and $\op J\rho = 0$.
\end{enumerate}
To this end, we let $\ell$ denote such an eigenvalue and $\rho$ the corresponding eigenoperator, which is normalized as $\norm{\rho}^2 = \tr(\rho^\dagger \rho) = 1$. We then have that
\begin{align}
	\op L_0 \rho &= \ell \rho
\shortintertext{and}
	\op L_0 \rho^\dagger &= -\ell \rho^\dagger .
\end{align}
A short calculation now shows that
\begin{align}
    	0 &= \tr( \rho^\dagger \op L_0 \rho ) + \tr( \rho\, \op L_0 \rho^\dagger ) \nonumber\\
		&= \Re\bigl[ \tr( \rho^\dagger \op L' \rho ) \bigr] - \sum_\mu \gamma_\mu \eta_\mu \bigl( \norm{L_\mu \rho}^2 + \norm{L_\mu \rho^\dagger}^2 \bigr).
\end{align}
Since every term on the right-hand side is non-positive, they must all be zero.
Therefore, for all channels, where~$\eta_\mu$ is non-zero, both $L_\mu \rho$ and $L_\mu \rho^\dagger$ must be zero, which proves our statement (3).

We now immediately see that if $\op L_0 \rho = 0$, then $\rho$ must be a dark state.
For statement (1), we still need to show that the converse holds true. To this end, we write the dark state $\rho$ as $\rho = X^\dagger X$ for some operator $X$, which is possible since $\rho$ is positive semi-definite.
We then find
\begin{equation}
	\tr(\op J \rho) = \sum_\mu \gamma_\mu \eta_\mu\, \norm{X L_\mu^\dagger}^2,
\end{equation}
and therefore $\op J \rho = 0$ implies $\acomm{D}{\rho} = 0$, and we are done.

Our statement (2) can be shown as in the case of a completely positive trace-preserving generator in lemma 17 of Ref.~\cite{BaumgartnerJPhysA2008a}. Using statement (3), we additionally find that:
\begin{enumerate}[(1), leftmargin=*, start=4]
\item If $\op L$ does not have purely imaginary eigenvalues, then $\op L_0$ does not have such eigenvalues either.
\item If $\op L_0$ has purely imaginary eigenvalues, the corresponding eigenoperators are traceless.
\end{enumerate}

\section{Oscillations in the Jump-Time Evolution} \label{app:oscill}

Here, we show that the jump-time evolution $\bar\rho_n$ can exhibit oscillations for large $n$ even if the wall-time evolution $\rho_t$ approaches a unique steady state.
To this end, we consider a two-state system whose time evolution is governed by a classical master equation with the rate matrix
\begin{equation}
	\op L = \begin{pmatrix}
		-1 & 1 \\ 1 & -1
	\end{pmatrix} .
\end{equation}
Counting all jumps between the states, the jump operator is
\begin{equation}
	\op J = \begin{pmatrix}
		0 & 1 \\ 1 & 0
	\end{pmatrix} .
\end{equation}

Using Eqs.~\eqref{eq:fnt} and \eqref{eq:convolution2}, we easily determine
\begin{equation}
	\op F_1 = \begin{pmatrix}
		0 & 1 \\ 1 & 0
	\end{pmatrix} ,
\end{equation}
which swaps the two entries of the jump-time average state $\bar\rho_n$.
Therefore, the jump-time evolution shows oscillatory behavior: $\bar\rho_n = \rho_0$ for even $n$ and $\bar\rho_n = \op F_1 \rho_0$ for odd $n$.
However, the wall-time averaged state approaches the steady state $\rho_\infty = (1/2, 1/2)$.

\section{Information Content of First-Passage Propagators} \label{app:quick_example}

Here, we show that different system dynamics can lead to the same first-passage propagator. To this end, we
consider the three-state system illustrated in Fig.~\ref{fig:appc}(a).
Its time evolution is governed by a classical master equation with the rate matrix
\begin{equation}
	\op L = \begin{pmatrix}
		-\gamma - \alpha & \alpha & \gamma \\
		\alpha & -\gamma - \alpha & 0 \\
		\gamma & \gamma & -\gamma
	\end{pmatrix} .
\end{equation}
The system switches with the rate $\alpha$ between two excited states and eventually decays into the ground state.
The decay rate $\gamma$ is the same for both excited states.
The switching rate is therefore irrelevant for the timing of the decay into the ground state.

From the ground state, the system eventually transitions back into the excited state. To count these jumps, we define the jump operator as
\begin{equation}
	\op J = \begin{pmatrix}
		0 & 0 & \gamma \\
		0 & 0 & 0 \\
		0 & 0 & 0
	\end{pmatrix} .
\end{equation}
The first-passage propagator then becomes
\begin{equation}
	\op F_1(s) = \begin{pmatrix}
		(1 + s/\gamma)^{-2} & (1 + s/\gamma)^{-2} & (1 + s/\gamma)^{-1} \\
		0 & 0 & 0 \\
		0 & 0 & 0
	\end{pmatrix},
\end{equation}
which is independent of the switching rate $\alpha$.
Hence, we see that different system dynamics can lead to the same first-passage propagator.

In agreement with our discussion in Sec.~\ref{subsec:main_thm}, the loss of information in the first-passage propagator is accompanied by a counting-field independent eigenvalue of the tilted generator $\op L(z)$.
This eigenvalue, which is $-\gamma - 2\alpha$, cannot be inferred from the eigendecomposition of $\op F_1(s)$.

Physically, the loss of information happens since both the first-passage time distribution and the jump-time averaged states do not depend on the switching rate $\alpha$.
Figures~\ref{fig:appc}(b-c) show modifications of our example that illustrate this statement.
In Fig.~\ref{fig:appc}(b), the decay rates from the excited states into the ground state differ.
The rate matrix and the jump operator thus become
\begin{align}
	\op L = \begin{pmatrix}
		-\gamma_{31} - \alpha & \alpha & \gamma \\
		\alpha & -\gamma_{32} - \alpha & 0 \\
		\gamma_{31} & \gamma_{32} & -\gamma
	\end{pmatrix}
		\quad \text{and} \quad
	\op J = \begin{pmatrix}
		0 & 0 & \gamma \\
		0 & 0 & 0 \\
		0 & 0 & 0
	\end{pmatrix}
\end{align}
with $\gamma_{31} \neq \gamma_{32}$.
The first-passage time distribution becomes $\alpha$-dependent, while the jump-time averaged states are still $\bar\rho_n = (1, 0, 0)$, which is independent of $\alpha$.
In Fig.~\ref{fig:appc}(c), the ground state is split into two states to keep track of which excited state the trajectory came from, and the system jumps back from a ground state into the original excited state.
The rate matrix and jump operator are
\begin{align}
	\op L = \begin{pmatrix}
		-\gamma - \alpha & \alpha & \gamma & 0 \\
		\alpha & -\gamma - \alpha & 0 & \gamma \\
		\gamma & 0 & -\gamma & 0 \\
		0 & \gamma & 0 & -\gamma
	\end{pmatrix}
		\quad \text{and} \quad
	\op J = \begin{pmatrix}
		0 & 0 & \gamma & 0 \\
		0 & 0 & 0 & \gamma \\
		0 & 0 & 0 & 0 \\
		0 & 0 & 0 & 0
	\end{pmatrix} .
\end{align}
Here, the first-passage time distribution is the same as in the original example, but the jump-time averaged states become $\alpha$-dependent.
In both modifications, the first-passage propagator becomes $\alpha$-dependent, and the tilted generator does not have a counting-field independent eigenvalue, meaning that it is fully determined by the first-passage propagator.

\section{Left Eigenstates} \label{app:steadystates}

We here consider a system with multiple steady states.
The eigenspaces of the generators $\op L = \op L(z=1)$ and $\op F_1 = \op F_1(s=0)$ for the eigenvalues zero and one are therefore degenerate.
Our goal is to show that bases $\rho_i$ and $\bar\rho_i \propto \op J \rho_i$ of these eigenspaces exist such that the dual bases of the space of left zero eigenstates are the same.

Generically, the degeneracy of the eigenspaces is lifted away from $z=1$ and $s=0$.
There are therefore unique states $\rho_i(z)$ and functionals $\varphi_i(z)$ such that
\begin{align}
	\op L(z) \rho_i(z) &= \ell_i(z) \rho_i(z)
\shortintertext{and}
	\varphi_i(z) \op L(z) &= \ell_i(z) \varphi_i(z)
\end{align}
with $\lim_{z \to 1} \ell_i(z) = 0$, normalized such that $\tr \rho_i(z) = 1$ and $\varphi_j(z) \rho_i(z) = \delta_{ij}$.
We claim that the basis
\begin{equation}
	\rho_i = \lim_{z \to 1} \rho_i(z)
\end{equation}
has the desired property; that is, this basis is the ``particular basis'' from the main text.
We also conjecture that these basis states are always physical states.

To prove our claim, we first note that the corresponding dual basis is given by $\varphi_i = \lim_{z \to 1} \varphi_i(z)$, because
\begin{equation}
	\varphi_j \rho_i = \lim_{z \to 1} \bigl[ \varphi_j(z) \rho_i(z) \bigr] = \delta_{ij} .
\end{equation}
By the fourth statement of our correspondence, $\op F_1(\ell_i(z))$ has the eigenvalue $z$, and the left and right eigenstates are
\begin{align}
	\bar\varphi_i(z) &= \mathcal N_1(z)\, \varphi_i(z)
\shortintertext{and}
	\bar\rho_i(z) &= \mathcal N_2(z)\, \op T_{\ell_i(z)}(0)^{-1} \rho_i(z) .
\end{align}
Here, the normalization constants $\mathcal N_1(z)$ and $\mathcal N_2(z)$ are chosen such that $\tr \bar\rho_i(z) = 1$ and $\bar\varphi_j(z) \bar\rho_i(z) = \delta_{ij}$.
Clearly, $\bar\rho_i = \lim_{z\to 1} \bar\rho_i(z) \propto \op J \rho_i$, and the corresponding dual basis is $\bar\varphi_i = \lim_{z \to 1} \bar\varphi_i(z)$.
We have thus shown that $\varphi_i$ and $\bar\varphi_i$ are proportional for all $i$.
We still need to prove that they are equal.
It suffices to show that $\varphi_i \bar\rho_i = 1$.

We first note that $\op F_1^n \rho_i$ must converge to a jump-time steady state for large $n$.
Since $(\bar\varphi_j \op F_1^n) \rho_i = \bar\varphi_j \rho_i = 0$ for $i \neq j$, it can only be
\begin{equation}
	\lim_{n \to \infty} \op F_1^n\, \rho_i = \bar\rho_i .
\end{equation}
We thus obtain $\varphi_i \bar\rho_i = \lim_{n \to \infty} \varphi_i \op F_1^n\, \rho_i = \varphi_i\rho_i = 1$, which completes our proof.

\section{Propagators from Keldysh's Theorem} \label{app:keldysh}

For  dynamics with forward and backward jumps, our goal is to express the $n$-resolved propagators \eqref{eq:Ttn_general} in terms of eigenvalues and eigenvectors of the tilted generator
\begin{equation}
	\op L(z) = \op L_0 + z^{-1} \op J_+ + z \op J_-.
\end{equation}
The operator $\op L(z)$ generates the time evolution with the counting field $z$ included, so that
\begin{equation} \label{eq:keldysh:fcs}
	\op T_t(z) = \ee^{\op L(z) t},
\end{equation}
where 
\begin{equation}
	\op T_t(z) = \sum_n z^{-n} \op T_t(n)
\end{equation}
is the discrete Laplace transform of the $n$-resolved propagators, $\op T_t(n)$.

From Eq.~\eqref{eq:keldysh:fcs}, the $n$-resolved propagators can be determined by contour integration.
By also performing a Laplace transform in time, we find
\begin{equation} \label{eq:keldysh:contour}
	\op T_s(n) = \int_0^\infty \dd t\, \ee^{-st} \oint_\gamma \frac{\dd z}{2\pi\ii} z^{n-1}\, \ee^{\op L(z) t},
\end{equation}
where $\gamma$ is any contour that encircles the origin once. If we could exchange the integrals, we would get~\cite{RudgeJChemPhys2019}
\begin{equation} \label{eq:keldysh:pencil_integral}
	\op T_s(n) = \oint_\gamma \frac{\dd z}{2\pi\ii} \frac{z^{n-1}}{s - \op L(z)}.
\end{equation}
However, since the integrand has poles that are different from $z=0$, this expression now depends on the contour.
Specifically, the integrand in Eq.~\eqref{eq:keldysh:pencil_integral} has poles at those points $z$, where $s$ -- considered here as a parameter -- is an eigenvalue of $\op L(z)$.
We denote these poles by $z_i(s)$, where the index $i$ enumerates different poles.
These poles are solutions to the equations
\begin{equation}
	\ell_j(z) = s ,
\end{equation}
where the index $j$ enumerates the eigenvalues $\ell_j(z)$ of $\op L(z)$.
We choose the functions $z_i(s)$ to be holomorphic, apart from some singularities and branch cuts, a choice that does not hinder the following arguments.

Since $\op L(z)$ diverges only at $z=0$ and $z\rightarrow\infty$, the poles $z_i(s)$ must move towards either the origin or infinity as the real part of $s$ grows large.
For any contour $\gamma$, the integrals in Eq.~\eqref{eq:keldysh:contour} can be exchanged if the real part is large enough.
Therefore, Eq.~\eqref{eq:keldysh:pencil_integral} is correct, if the contour contains only the poles $z_i(s)$ that move towards the origin.
Thus, for all $s$, we set
\begin{equation} \label{eq:keldysh:residues}
	\op T_s(n) = \Res_{0}\biggl[ \frac{z^{n-1}}{s - \op L(z)} \biggr] + \sum_{\mathclap{\substack{i\\z_i \to 0}}} \Res_{z_i(s)}\biggl[ \frac{z^{n-1}}{s - \op L(z)} \biggr],
\end{equation}
where $\Res_z$ denotes the residue at the point $z$, and the sum only includes those poles with $z_i(s) \to 0$ as $\Re(s) \to \infty$.
Equation \eqref{eq:keldysh:residues} constitutes a continuation of the conditionally defined integral \eqref{eq:keldysh:contour} to the complex plane, just as we set $\op T_s(0) = (s - \op L_0)^{-1}$ for all $s$ in Eq.~\eqref{eq:convolution2}.

Equation \eqref{eq:keldysh:residues} is inconvenient for negative $n$ because of the higher-order poles that are involved. Substituting instead $z \to 1/z$ in Eq.~\eqref{eq:keldysh:contour} and following the same arguments, we find the equivalent formula
\begin{equation}
	\op T_s(n) = \Res_0\biggl[ \frac{z^{-n-1}}{s - \op L(1/z)} \biggr] + \sum_{\mathclap{\substack{i\\z_i \to \infty}}} \Res_{1/z_i(s)}\biggl[ \frac{z^{-n-1}}{s - \op L(1/z)} \biggr]
\end{equation}
for $n \leq 0$, summing now over the poles with $z_i(s) \to \infty$ as $\Re(s) \to \infty$.

Finally, we evaluate the residues in these expressions using Keldysh's theorem~\cite{KeldyshRussMathSurv1971, Mennicken2003, BeynLinearAlgebraItsAppl2012}.
The points $z = 0$ and $z = z_i(s)$ are eigenvalues of the operator pencil
\begin{equation}
	\op A(z) = zs - z\op L(z) .
\end{equation}
For $z=0$, the left and right eigenstates $\varphi_{0k}$ and $\rho_{0k}$ of the operator pencil are the left and right eigenstates of $\op J_+$ for the eigenvalue zero. For $z = z_i(s)$, its left and right eigenstates $\varphi_{ik}(s)$ and $\rho_{ik}(s)$ are the left and right eigenstates of $\op L(z_i(s))$ for the eigenvalue $s$.

For the sake of simplicity, we assume these eigenvalues to be semi-simple, which means that no higher-order poles occur. For eigenvalues that are not semi-simple, generalized eigenstates, similar to those appearing in the Jordan normal form, would need to be considered in addition to the eigenstates listed above. We note that $z=0$ can be a semi-simple eigenvalue according to the theory of operator pencils, even if $\op J_+$ is not diagonalizable.

For semi-simple eigenvalues, the left and right eigenstates can be normalized such that
\begin{align}
	&\varphi_{0k}\, \op A'(0)\, \rho_{0k'}(s) = \delta_{k,k'}
\shortintertext{and}
	&\varphi_{ik}(s)\, \op A'(z_i(s))\, \rho_{ik'}(s) = \delta_{k,k'} ,
\end{align}
where the prime denotes the derivative with respect to $z$.
We can then apply Keldysh's theorem to obtain the final result
\begin{equation}
	\op T_s(n) = \delta_{n,0} \sum_k \rho_{0k} \varphi_{0k} + \sum_{\mathclap{\substack{i\\z_i \to 0}}} \sum_k z_i(s)^n \rho_{ik}(s) \varphi_{ik}(s)
\end{equation}
for $n \geq 0$.

For $n \leq 0$, we consider the operator pencil
\begin{equation}
	\op B(z) = zs - z\op L(1/z).
\end{equation}
For $z=0$, the left and right eigenstates $\bar\varphi_{0k}$ and $\bar\rho_{0k}$ of this operator pencil are the left and right eigenstates of $\op J_-$ for the eigenvalue zero, and for $z = 1/z_i(s)$, its left and right eigenstates $\bar\varphi_{ik}(s)$ and $\bar\rho_{ik}(s)$ are the left and right eigenstates of $\op L(z_i(s))$ for the eigenvalue $s$.
The eigenstates must now be normalized so that
\begin{align}
	&\bar\varphi_{0k}\, \op B'(0)\, \bar\rho_{0k'} = \delta_{k,k'}
\shortintertext{and}
	&\bar\varphi_{ik}(s)\, \op B'(1/z_i(s))\, \bar\rho_{ik'}(s) = \delta_{k,k'} .
\end{align}
The final result for $n \leq 0$ is then
\begin{equation}
	\op T_s(n) = \delta_{n,0} \sum_k \bar\rho_{0k} \bar\varphi_{0k} + \sum_{\mathclap{\substack{i\\z_i \to \infty}}} \sum_k z_i(s)^n \bar\rho_{ik}(s) \bar\varphi_{ik}(s) .
\end{equation}

\vbadness=10000
\hbadness=10000

\end{document}